\documentclass[preprint,10pt]{elsarticle}
\usepackage{setspace}

\usepackage{setspace}
\usepackage{amssymb} 
\usepackage{amsmath}
\usepackage{mathrsfs}
\usepackage{stmaryrd}
\usepackage{amsthm}
\usepackage{graphicx}
\usepackage{comment}
\usepackage{longtable}
\usepackage[colorinlistoftodos]{todonotes}
\usepackage{tikz-cd}   
\usepackage[utf8]{inputenc} 
\usepackage[T1]{fontenc}    
\usepackage{hyperref}       
\usepackage{url}            
\usepackage{booktabs}       
\usepackage{amsfonts}       
\usepackage{nicefrac}       
\usepackage{microtype}      
\usepackage{amsfonts}       
\usepackage{nicefrac}       
\usepackage{lingmacros}
\usepackage{tree-dvips} 
\usepackage{nccmath}
\usepackage{mathtools}
\usepackage{bbm}
\usepackage{graphicx}  

\DeclareMathOperator*{\argmax}{argmax}
\usepackage{amsmath, amssymb, mathtools, bm, etoolbox}
\usepackage[noend]{algpseudocode}
\usepackage{algorithm}
\usepackage {mathrsfs}
\usepackage{caption}
\usepackage{graphicx}
\usepackage{subcaption}
\usepackage{color}
\usepackage{enumitem}

\newtheorem{thm-defn}[theorem]{Theorem/Definition}

\theoremstyle{definition}

\theoremstyle{remark}

\newcommand{\ignore}[1]{}{}
\usepackage{graphicx}
\usepackage{amssymb}

\begin{document}

\begin{frontmatter}


\title{Dynamics, behaviours, and anomaly persistence in cryptocurrencies and equities surrounding COVID-19}
\author{Nick James} \ead{nicholas.james@sydney.edu.au}
\address{School of Mathematics and Statistics, University of Sydney, NSW, Australia}

\begin{abstract}
This paper uses new and recently introduced methodologies to study the similarity in the dynamics and behaviours of cryptocurrencies and equities surrounding the COVID-19 pandemic. We study two collections; 45 cryptocurrencies and 72 equities, both independently and in conjunction. First, we examine the evolution of cryptocurrency and equity market dynamics, with a particular focus on their change during the COVID-19 pandemic. We demonstrate markedly more similar dynamics during times of crisis. Next, we apply recently introduced methods to contrast trajectories, erratic behaviours, and extreme values among the two multivariate time series. Finally, we introduce a new framework for determining the persistence of market anomalies over time. Surprisingly, we find that although cryptocurrencies exhibit stronger collective dynamics and correlation in all market conditions, equities behave more similarly in their trajectories and extremes, and show greater persistence in anomalies over time.


\end{abstract}


\end{frontmatter}


\section{Introduction}
\label{Introduction}

Over the last several years there has been growing interest in the cryptocurrency market. The sector has experienced impressive growth in asset inflows and its level of sophistication. More recently, the COVID-19 pandemic has caused immense social and economic impacts, including changes in the behaviour of financial markets. The goal of this paper is to analyze the evolution of cryptocurrencies and equities over time, and in particular, assess whether the increase in interest from sophisticated investors has led to more uniformity in the dynamics and behaviours of the two asset classes. We use the COVID-19 pandemic as a motivating example to ascertain whether this similarity changes during market crises.

The study of financial market correlation structure has been a topic of great interest to the nonlinear dynamics community over the past several decades \cite{Fenn2011, Laloux1999, Mnnix2012}. Evolutionary market dynamics have been studied through a wide variety of techniques such as clustering \cite{Heckens2020} and principal components analysis (PCA) \cite{Laloux1999, Kim2005, Pan2007, Wilcox2007}. Until the past decade, the primary asset classes of interest to the research community were equities \cite{Wilcox2007}, fixed income \cite{Driessen2003}, and foreign exchange \cite{Ausloos2000}. More recently, select research has focused on the study of trajectory modelling \cite{James2021}, extreme behaviours and structural breaks \cite{James2021_crypto, Telli2020} and these methods have been applied to a variety of asset classes.

There is a current wave of interest from econophysics researchers in the development and application of methods for understanding cryptocurrency dynamics. Areas attracting interest from researchers include studies of Bitcoin's behaviour \cite{Chu2015, Lahmiri2018, Kondor2014, Bariviera2017, AlvarezRamirez2018}, fractal patterns, \cite{Stosic2019, Stosic2019_2, Manavi2020, Ferreira2020}, cross-correlation and scaling effects \cite{Drod2018, Drod2019, Drod2020, Gbarowski2019, Wtorek2020}. Many of these studies are concerned with the time-varying nature of such dynamics, or the behaviour of cryptocurrencies during various market regimes. Quite naturally, the impact of COVID-19 on cryptocurrency behaviours has been widely studied \cite{Corbet2020, Conlon2020, Conlon2020_2, Ji2020}

The evolution of COVID-19 and its impact on financial markets has attracted broad interest from various research communities. COVID-19's spread and containment measures have been studied by the epidemiology community \cite{Wang2020, Chinazzi2020, Liu2020, Fang2020, Zhou2020, Dehning2020, Ferguson, Wang2020_chaos, Maier2020}, while clinically-inclined research has detailed new treatments for various COVID-19 strains \cite{Jiang2020, Zu2020, Li2020, Zhang2020, Wang2020_CELL, NEJM2020, Corey2020}. The pandemic's varied impact on financial markets has also been studied \cite{Zhang2020finance, He2020, Zaremba2020}, with many papers exploring financial contagion and market stability \cite{Akhtaruzzaman2020, Okorie2020, Lahmiri2020}. In the nonlinear dynamics community, COVID-19 research has used new and existing techniques to study the temporal evolution of cases and deaths \cite{Khajanchi2020, Ribeiro2020, Chakraborty2020,James2020_chaos}, with a substantial emphasis on SIR models \cite{SIRBallesteros2020,SIRBarlow2020,SIRCadoni2020,SIRComunian2020,SIRNeves2020,SIRVyasarayani2020,SIRWeinstein2020}, power law models \cite{Beare2020,Manchein2020,Blasius2020,Anastassopoulou2020} and the use of networks \cite{Thurner2020}. More recently there has been work that explores the impact of COVID-19 cases on the performance of country financial markets \cite{James2021}.

The goal of this paper is to explore the similarity in the dynamics and behaviours of cryptocurrencies and equities over the past two years. In doing so, we make several contributions. First, we complement current methods with the introduction of a new measure between eigenspectra to study the similarity in two time series' evolutionary dynamics. Next, we apply recently developed techniques to study the trajectories, extremes and erratic behaviour of cryptocurrencies and equities, and analyze their similarity. Finally, we introduce a pithy method to study the persistence of financial anomalies over time. 

This paper is structured as follows. Section \ref{data} describes the data used in this paper. Section \ref{Market_dynamics} studies the time-varying dynamics of cryptocurrencies and equities, and contrasts their behaviour during different market states. Sections \ref{trajectory_modelling}, \ref{erratic_behaviour} and  \ref{Extreme_behaviour} study trajectories, erratic behaviour and extremes respectively. In Section \ref{anomaly persistence} we contrast the consistency in anomalies among our two collections. In Section \ref{Conclusion}, we conclude.

\section{Data}
\label{data}
In the proceeding analysis, the two primary objects of study are cryptocurrency and equity multivariate time series between 03-12-2018 to 08-12-2020. We analyze the 45 largest cryptocurrencies by market capitalisation (excluding those previously identified as anomalous) \cite{James2021_crypto} and 72 global equities whose market capitalisation is greater than US\$100 billion. We report and contrast on the dynamics, behaviours, and anomaly persistence between cryptocurrencies and equities. In select sub-sections, we refer to the period 03-12-2018 to 28-02-2020 as Pre-COVID, 02-03-2020 to 29-05-2020 as Peak COVID, and 01-06-2020 to 08-12-2020 as Post-COVID. Cryptocurrency and equity data are sourced from https://coinmarketcap.com/ and Bloomberg respectively. A full list of cryptocurrencies and equities studied in this paper is available in \ref{appendix:mathematical_objects}.

\section{Market dynamics}
\label{Market_dynamics}

\subsection{Evolutionary dynamics}
\label{evolutionary_dynamics}
In this section we follow the framework introduced in \cite{Fenn2011} to study the temporal evolution of correlation structure in cryptocurrencies and equities, and contrast these collections' evolutionary dynamics. Our analysis in this section differs from \cite{Fenn2011} in several ways, however. First, rather than applying this framework to a single collection of securities from different asset classes, we apply the time-evolving model to two separate asset classes (cryptocurrencies and equities) and compare the respective time-varying dynamics. Second, we use a shorter time window to study correlation structures, allowing correlations to change more quickly to varying market conditions. This allows us to study the impact of COVID-19 on both collections. Third, we introduce a new \emph{dynamics deviation} measure between surfaces to determine the similarity in two time-varying eigenspectra across different time periods. Finally, we use daily data rather than weekly data.

Let $c_i(t)$ and $e_j(t)$ be the multivariate time series of cryptocurrency and equity daily closing prices, for $t=1,...,T$, $i=1,..., N$, and $j=1,...,K$. We generate two multivariate time series of log returns, $R^c_i(t)$ and $R^e_j(t)$, where cryptocurrency and equity log returns are computed as follows

\begin{align}
R^{c}_{i}{(t)} &= \log \left(\frac{c_i{(t)}}{c_i{(t-1)}}\right), \\
R^{e}_{j}{(t)} &= \log \left(\frac{e_j{(t)}}{e_j{(t-1)}}\right).
\end{align}

We define standardized cryptocurrency returns as $\hat{R}^c_i(t) = [R^c_i(t) - \langle R^c_i \rangle] / \sigma(R^c_i)$, where $\sigma(R^c_i) = \sqrt{\langle (R^{c}_i)^2 \rangle - \langle R^c_i \rangle ^ 2}$ represents the standard deviation of cryptocurrency time series $R_i^c$ and $\langle . \rangle $ denotes an average over time. Standardized equity returns are computed similarly and we denote this time series  $\hat{R}^e_j(t)$. Having normalized the returns, we may construct empirical correlation matrices

\begin{align}
\Omega^c = \frac{1}{T_1} \hat{R^c} \hat{R^c}^T, \\
\Omega^e = \frac{1}{T_1} \hat{R^e} \hat{R^e}^T,
\end{align}

for cryptocurrency and equity time series. Elements of both correlation matrices $\omega^c(i,j)$ and $\omega^e(i,j)$ lie $\in [-1,1]$. We study the evolution of these correlation matrices, using a rolling window of $T_1=60$ days. One must be judicious in the choice of the smoothing parameter $T_1$, as correlation coefficients can be excessively smooth or noisy if $T_1$ is too large or too small respectively. Our choice of 60 days corresponds approximately to the length of the COVID-19 market crash. This allows us to capture the entirety of the COVID-19 market shock, without including unrelated data outside the COVID-related crash in our calculation. Next, we study the dynamics of the cryptocurrency and equity market security collections by applying principal components analysis (PCA) to the two time-varying correlation matrices. For each correlation matrix, we wish to estimate the linear maps $\Phi^c$ and $\Phi^e$ that transform our standardized cryptocurrency returns $\hat{R}^c$ and equity returns $\hat{R}^e$ into uncorrelated variables $Z^c$ and $Z^e$ respectively. That is,
\begin{align}
    Z^c &= \Phi^c \hat{R}^c, \\
    Z^e &= \Phi^e \hat{R}^e.
\end{align}
where the rows of $Z^c$ and $Z^e$ represent PCs of the matrices $R^c$ and $R^e$. The rows of $\Phi^c$ and $\Phi^e$, which contain PC coefficients, are ordered such that the first rows are along the axes of most variation in the data, with subsequent PCs, subject to the optimization constraint that they are all mutually orthogonal, each accounting for maximal variance along their respective axes. The correlation matrices, which are symmetric and diagonalizable matrices can be written in the form 
\begin{align}
    \Omega^c = \frac{1}{T_1} \Lambda_c D^c \Lambda_c^{T} , \\
    \Omega^e = \frac{1}{T_1} \Lambda_e D^e \Lambda_e^{T},
\end{align}
where $D^c$, $D^e$ are diagonal matrices with eigenvalues $\lambda^c_i$, $\lambda^e_j$, and $\Lambda_c$, $\Lambda_e$ are orthogonal matrices with the associated eigenvectors from the cryptocurrency and equity correlation matrices respectively. The PCs are estimated using the diagonalizations above. 

Finally, we contrast the proportion of variance produced by a sub-collection of eigenvalues within each of the two collections. The total variance of the cryptocurrency returns $\hat{R}^c$ and equity returns $\hat{R}^e$ for the $N$ and $K$ assets respectively, is equal to the sum of all eigenvalues $\lambda^c_1+...+\lambda^c_N$ and $\lambda^e_1+....+\lambda^e_K$. This is equivalently the trace of the two diagonal matrices of eigenvalues $\text{tr}(D^c)=N$ and $\text{tr}(D^e)=K$. To compute the proportion of total variance explained by the $m^{th}$ PC in $\hat{R}^c$ and $\hat{R}^e$ is therefore $\Tilde{\lambda}_m^c =\lambda^c_m / N$ and $\Tilde{\lambda}_m^e = \lambda^e_m / K$. For more details on this construction readers should visit \cite{Fenn2011,Jolliffe2016, Jolliffe2011}. In some cases, the correlation matrix under examination may not be full rank. This is not a concern for the proceeding analysis, where we focus on the behavior of the first 10 eigenvectors. Since PCs are mutually orthogonal, and by definition linearly independent, this would not cause any issues in the conclusions resulting from our methodology.

\begin{figure*}
    \centering
    \begin{subfigure}[b]{0.48\textwidth}
        \includegraphics[width=\textwidth]{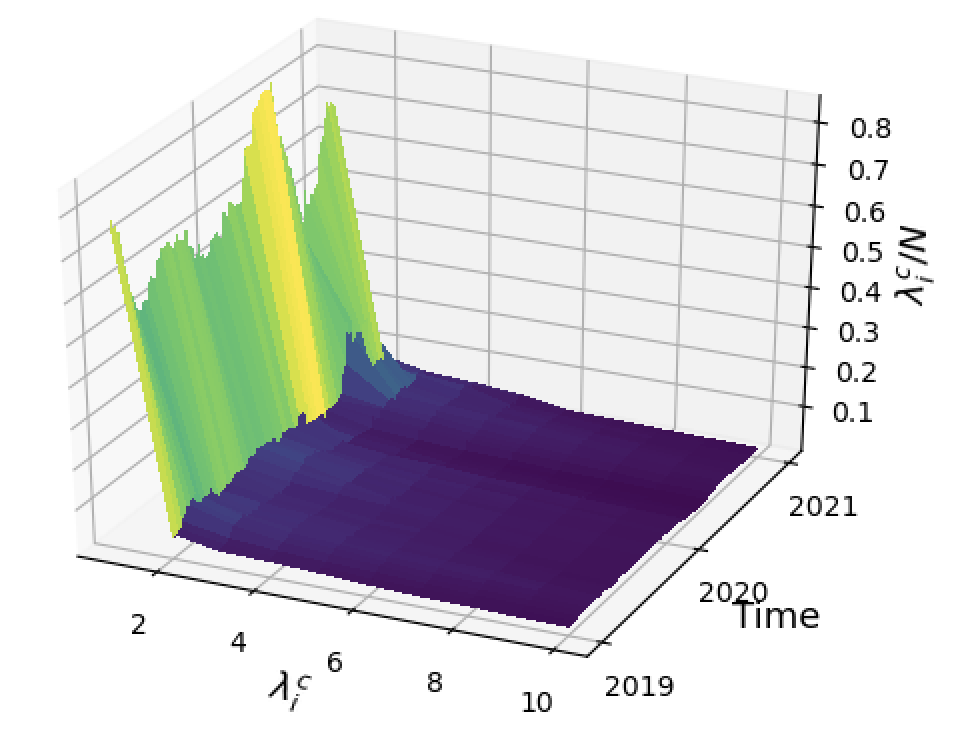}
        \caption{Cryptocurrency}
    \label{fig:Crypto_eigenspectrum}
    \end{subfigure}
    \begin{subfigure}[b]{0.48\textwidth}
        \includegraphics[width=\textwidth]{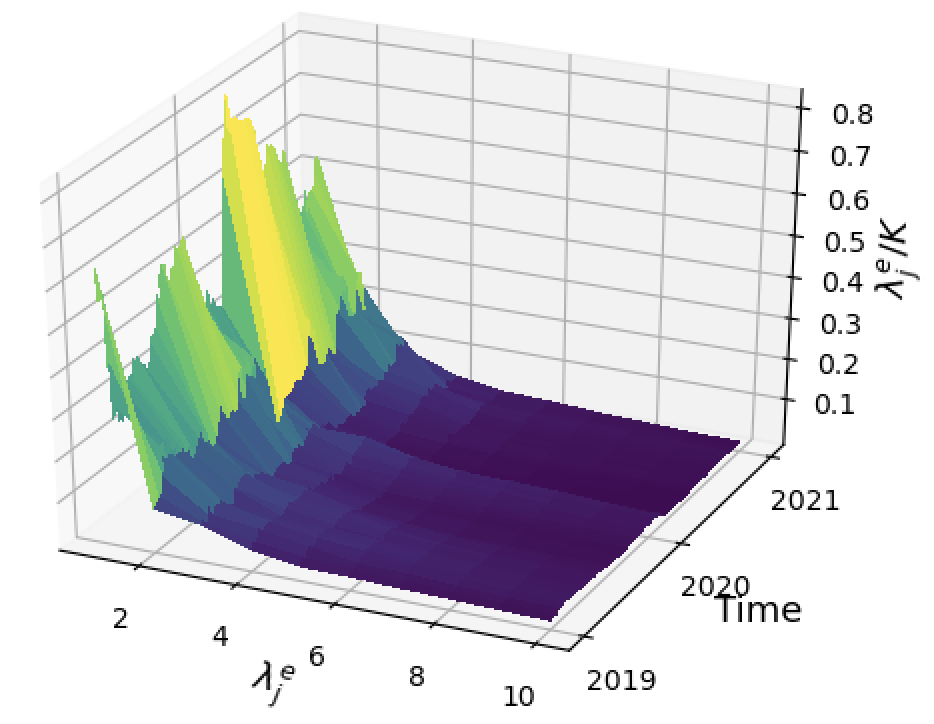}
        \caption{Equity}
    \label{fig:Equity_eigenspectrum}
    \end{subfigure}
        \caption{Time-varying eigenspectrum for cryptocurrencies and equities.}
    \label{fig:Eigenspectra}
\end{figure*}    

\begin{figure}
    \centering
    \includegraphics[width=0.85\textwidth]{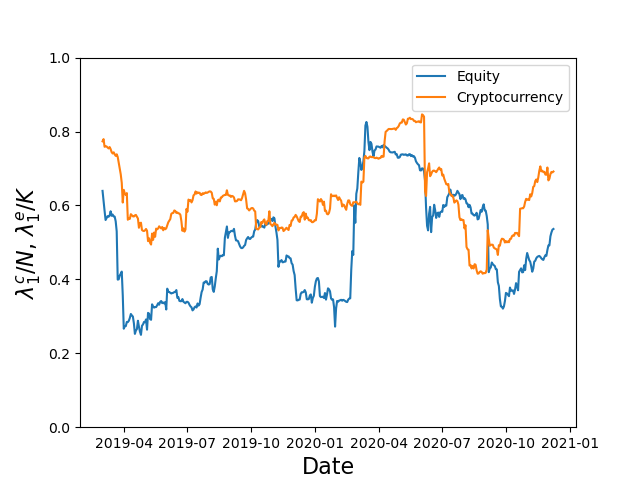}
    \caption{Rolling explained variance ratio for cryptocurrencies $\lambda^c_1/N$ and equities $\lambda^e_1/K$.}
    \label{fig:ExplainedVarianceRatio}
\end{figure}

\begin{figure*}
    \centering
    \begin{subfigure}[b]{0.48\textwidth}
        \includegraphics[width=\textwidth]{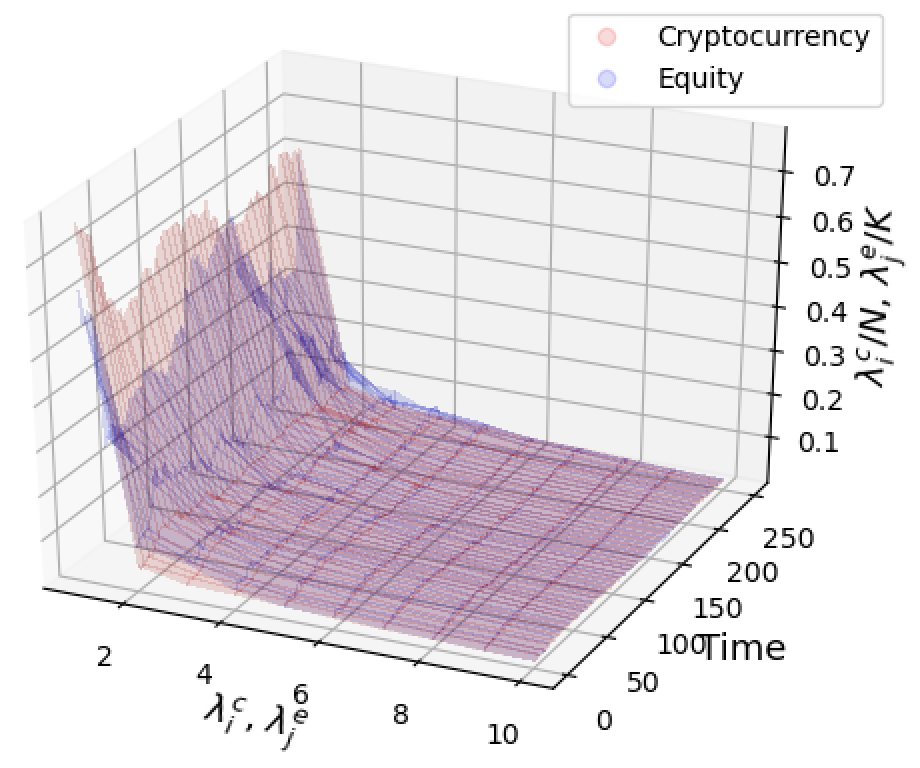}
        \caption{Pre Covid}
    \label{fig:pre_eigenspectrum}
    \end{subfigure}
    \begin{subfigure}[b]{0.48\textwidth}
        \includegraphics[width=\textwidth]{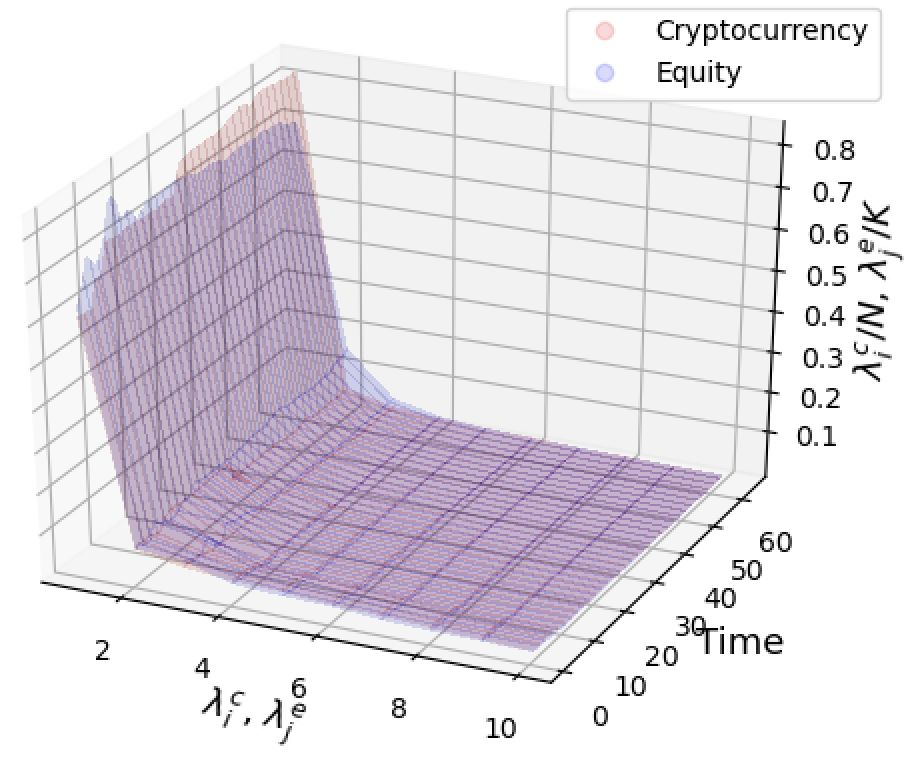}
        \caption{Peak Covid}
    \label{fig:peak_eigenspectrum}
    \end{subfigure}
    \begin{subfigure}[b]{0.48\textwidth}
        \includegraphics[width=\textwidth]{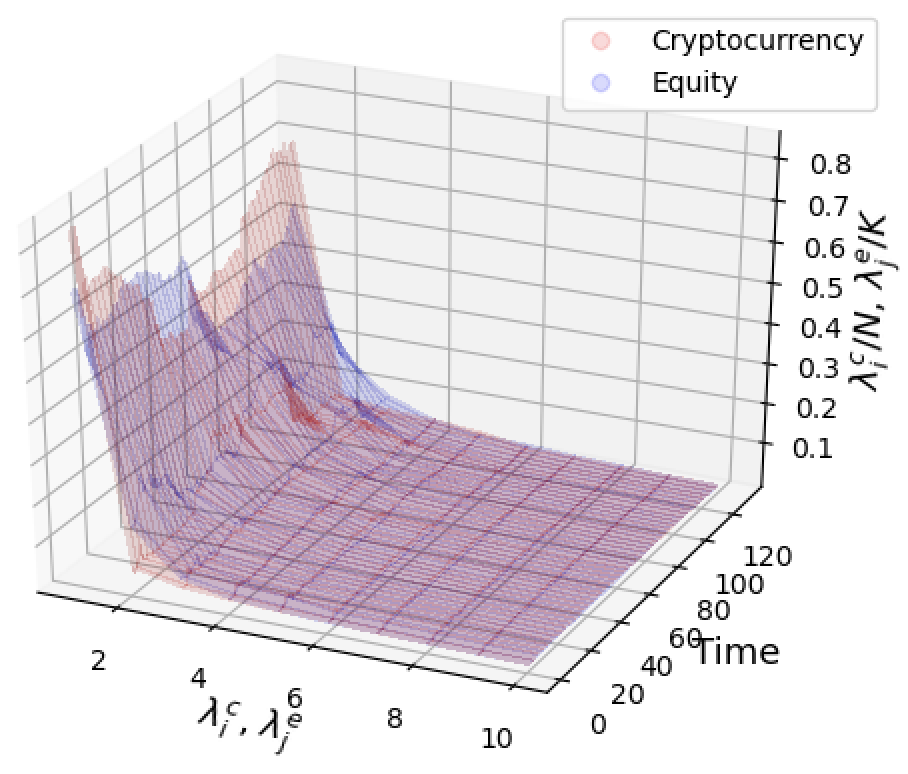}
        \caption{Post Covid}
    \label{fig:post_eigenspectrum}
    \end{subfigure}
        \caption{Time-varying eigenspectrum Pre Covid, Peak Covid, and Post Covid.}
    \label{fig:DiscreteEigenspectra}
\end{figure*}

Prior work has demonstrated that the eigenvector corresponding to the largest eigenvalue represents the significance of `the market' within the collection \cite{Fenn2011}. Bearing this in mind, there are several noteworthy insights revealed in Figure \ref{fig:Eigenspectra} regarding the similarity in cryptocurrency and equity dynamics. First, both Figure \ref{fig:Crypto_eigenspectrum} and Figure \ref{fig:Equity_eigenspectrum} exhibit a broadly similar shape; the majority of explanatory variance is provided by the first several eigenvectors, with the remaining proportion of total variance falling away quickly over the entire time period. Next, the explanatory variance provided by the first cryptocurrency eigenvalue $\Tilde{\lambda}_1^c$ seen in figure \ref{fig:Crypto_eigenspectrum} is consistently higher than the corresponding equity market eigenvalue $\Tilde{\lambda}^e_1$ in figure \ref{fig:Equity_eigenspectrum}. This demonstrates that the collective force of the market is more pronounced in cryptocurrencies than equities during our window of analysis. The second observation one may make from Figure \ref{fig:Eigenspectra} is the significant variability in $\Tilde{\lambda}^e_1$ when compared with $\Tilde{\lambda}_1^c$. This may highlight that within equity markets, there is more variability related to the market's impact on collective behaviour than that of cryptocurrencies. The time-varying explanatory variance of the first PC is displayed in Figure \ref{fig:ExplainedVarianceRatio}, where $\Tilde{\lambda}^c_1 > \Tilde{\lambda}^e_1$ for almost the entire window of analysis. Figure \ref{fig:ExplainedVarianceRatio} indicates that the difference in the first eigenvalue, $|\Tilde{\lambda}^c_1 - \Tilde{\lambda}^e_1|$, is smallest during the Peak COVID period.



\subsection{Dynamics surrounding COVID-19}
\label{static_dynamics}
In this section we study the similarity in the dynamics of cryptocurrencies and equities during three discrete time periods which are characterized by different systematic (market) risk profiles. Our goal is to determine whether the similarity in cryptocurrency and equity market dynamics changes in varying market conditions. The three periods are defined
\begin{itemize}
    \item Pre-COVID: 03-01-2018 to 28-02-2020,
    \item Peak COVID: 02-03-2020 to 29-05-2020,
    \item Post-COVID: 01-06-2020 to 08-12-2020,
\end{itemize}
with corresponding lengths $|T_{\text{PRE}}|$, $|T_{\text{PEAK}}|$ and $|T_{\text{POST}}|$. For the two sequences of time-varying correlation matrices $\Omega^c_t$ and $\Omega^e_t$, we study the similarity in the explanatory variance of the first 10 eigenvalues. $\Tilde{\lambda}^c_{1,t},...,\Tilde{\lambda}^c_{10,t}$ and $\Tilde{\lambda}^{e}_{1,t},...,\Tilde{\lambda}^{e}_{10,t}$. We define the difference in these spectral surfaces \emph{dynamics deviation} and compute as follows
\begin{align}
    \text{DD}_{\text{PRE}} &= \frac{1}{|T_{\text{PRE}}|} \sum^{311}_{t=60} \sum^{10}_{i=1} | \Tilde{\lambda}^c_{i,t} - \Tilde{\lambda}^e_{i,t} | \\
    \text{DD}_{\text{PEAK}} &=  \frac{1}{|T_{\text{PEAK}}|} \sum^{375}_{t=312} \sum^{10}_{i=1} | \Tilde{\lambda}^c_{i,t} - \Tilde{\lambda}^e_{i,t} | \\
    \text{DD}_{\text{POST}} &=  \frac{1}{|T_{\text{POST}}|} \sum^{508}_{t=376} \sum^{10}_{i=1} | \Tilde{\lambda}^c_{i,t} - \Tilde{\lambda}^e_{i,t}|.
\end{align}
The measure is normalized by the length of each time period, allowing us to compare dynamics during periods of varying lengths. As the majority of explanatory variance is provided by the first 10 eigenvalues in both the cryptocurrency and equity collections, we ignore the negligible difference in total variance explained by the remaining elements of the eigenspectrum.

\begin{table}[h]
\centering
\begin{tabular}{ |p{2.5cm}|p{2.5cm}|}
 \hline
 \multicolumn{2}{|c|}{Dynamics deviation scores} \\
 \hline
 Period & Score \\
 \hline
 $\text{DD}_{\text{PRE}}$ & 0.369 \\
 $\text{DD}_{\text{PEAK}}$ & 0.160 \\
 $\text{DD}_{\text{POST}}$ & 0.298 \\
\hline
\end{tabular}
\caption{Dynamics deviation from 3 periods of analysis}
\label{tab:table_dynamics_deviation}
\end{table}

\begin{figure*}
    \centering
    \begin{subfigure}[b]{0.48\textwidth}
        \includegraphics[width=\textwidth]{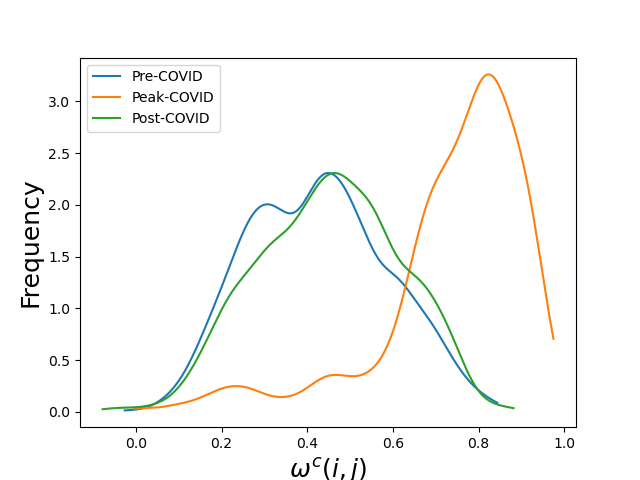}
        \caption{Cryptocurrency}
    \label{fig:Crypto_correlation}
    \end{subfigure}
    \begin{subfigure}[b]{0.48\textwidth}
        \includegraphics[width=\textwidth]{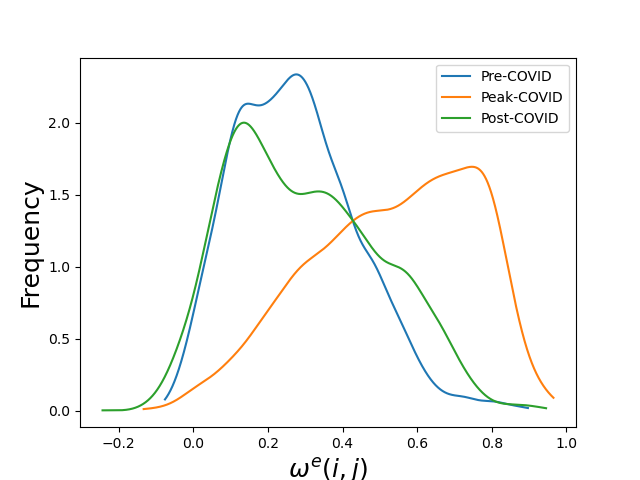}
        \caption{Equity}
    \label{fig:Equity_correlation}
    \end{subfigure}
        \caption{Kernel density estimates of $w^c(i,j)$ and $w^e(i,j)$ from Pre Covid, Peak Covid and Post Covid periods.}
    \label{fig:CorrelationCoefficients}
\end{figure*}

 Figure \ref{fig:DiscreteEigenspectra} shows the cryptocurrency and equity eigenspectra during the three windows of analysis. Of the three analysis windows, the two eigenspectra appear to be most similar during the Peak COVID period, which is displayed in Figure \ref{fig:peak_eigenspectrum}. This is confirmed in Table \ref{tab:table_dynamics_deviation}, which shows dynamics deviation scores for the three windows of analysis. The results highlight a significant increase in the similarity of the two collections' collective behaviour during the Peak COVID period, with a score of 0.160. The Pre-COVID and Post-COVID scores are 0.369 and 0.298 respectively, highlighting less similarity in the dynamics of equities and cryptocurrencies outside periods of market crisis. This is primarily due to the equity eigenspectrum's first eigenvalue exhibiting lower explanatory variance in the Pre-COVID and Post-COVID periods. This is shown in Figure \ref{fig:pre_eigenspectrum} and Figure \ref{fig:post_eigenspectrum} respectively.

Next, we contrast the cryptocurrency and equity correlation coefficients during the same three windows of analysis seen in Figure \ref{fig:CorrelationCoefficients}. This analysis is similar to that conducted in \cite{Drod2020}, where the probability density functions of off-diagonal correlation matrix elements are studied for six base cryptocurrencies and the USD. \cite{Drod2020}'s analysis also focuses on the distribution of correlation matrix eigenvalues, and the time-varying behavior of the correlation matrix's largest eigenvalues (seen in Figure 7 of \cite{Drod2020}). However, the work presented in this section focuses more prominently on changes in the distribution of correlation matrix elements seen in different markets across cryptocurrencies and equities. Figures \ref{fig:Crypto_correlation} and \ref{fig:Equity_correlation} display kernel density estimates of cryptocurrency and equity correlation matrix elements during the Pre-COVID, Peak COVID and Post-COVID periods. There are two important insights. First, both cryptocurrency and equity markets highlight a sharp increase in collective correlations during the Peak COVID period. Both cryptocurrencies and equities displayed markedly lower correlation coefficients in the Pre-COVID and Post-COVID periods. In all three periods, cryptocurrency correlation coefficients were more strongly positive than that of equities. These findings are consistent with the results in the Section \ref{evolutionary_dynamics}, where dynamics deviations were lowest during the Peak COVID market period; suggesting that during market crises cryptocurrency and equity behaviours are most similar.

\section{Trajectory modelling}
\label{trajectory_modelling}
In this section, we study the trajectory dynamics \cite{james2020covidusa} of equity and cryptocurrency closing prices for the entirety of our time period, a single period of $T = 508$ days. To compare trajectories of securities with markedly different prices, we normalize the cryptocurrency time series $c_i(t)$ and equity time series $e_j(t)$. Analyzing a candidate individual cryptocurrency provides a function $\mathbf{c}_{i} \in \mathbb{R}^{T}$. We let $\|\mathbf{c}_{i}\|_1 = \sum^{T}_{t=1} |c_{i}(t)|$ be the $L^1$ norm of the function, and define a normalized cryptocurrency price trajectory by $\mathbf{T}^{c}_{i} = \frac{\mathbf{c}_i}{\|\mathbf{c}_i\|_1}$. Similarly, we define $\|\mathbf{e}_{j}\|_1 = \sum^{T}_{t=1} |e_{j}(t)|$, and the corresponding normalized equity trajectory as $\mathbf{T}^{e}_{j} = \frac{\mathbf{e}_j}{\|\mathbf{e}_j\|_1}$. Distances between such vectors highlight the relative change in cryptocurrency and equity securities during the time period. To study such changes, we define two \emph{trajectory matrices}, $D^{TC}_{ij}=\|\mathbf{T}^c_i - \mathbf{T}^c_j\|_1$ and $D^{TE}_{ij} = \|\mathbf{T}^e_i - \mathbf{T}^e_j\|_1$, and perform \emph{hierarchical clustering}.

First, we compare norms of the two trajectory matrices to better understand similarity within each collection. Both of these matrices are symmetric, real and have trace 0. As the two collections are of different sizes, we normalize the norm computations by the number of elements in each trajectory matrix. The normalized cryptocurrency trajectory matrix norm $\|D^{TC}\|_{2^{*}} = N^{-1} \sqrt{\sum_{i,j} |d_{ij}^{tc}|^2}$ and the normalized equity trajectory matrix norm $\|D^{TE}\|_{2^{*}} = K^{-1} \sqrt{\sum_{i,j} |d_{ij}^{te}|^2}$. The normalized cryptocurrency trajectory matrix norm $\|D^{TC}\|_{2^{*}} = 0.4804$ and the normalized equity trajectory matrix norm $\|D^{TE}\|_{2^{*}} = 0.1849$, demonstrating more similarity in normalized trajectories among equities than cryptocurrencies. There are several possible explanations for this finding. First, there could be some bias in our sample of equities, having chosen the 72 largest equities in the world by market capitalization. It is likely that the volatility in their price behaviour will be less than that of smaller equities. On the other hand, this finding may quite reasonably reflect the volatile nature of cryptocurrency market sentiment. Although correlations among the cryptocurrency constituents are higher than that of equity constituents, the consistently strong influence of `the market' may make trajectories highly responsive to sharp changes in sentiment.

\begin{figure}
    \centering
    \includegraphics[width=\textwidth]{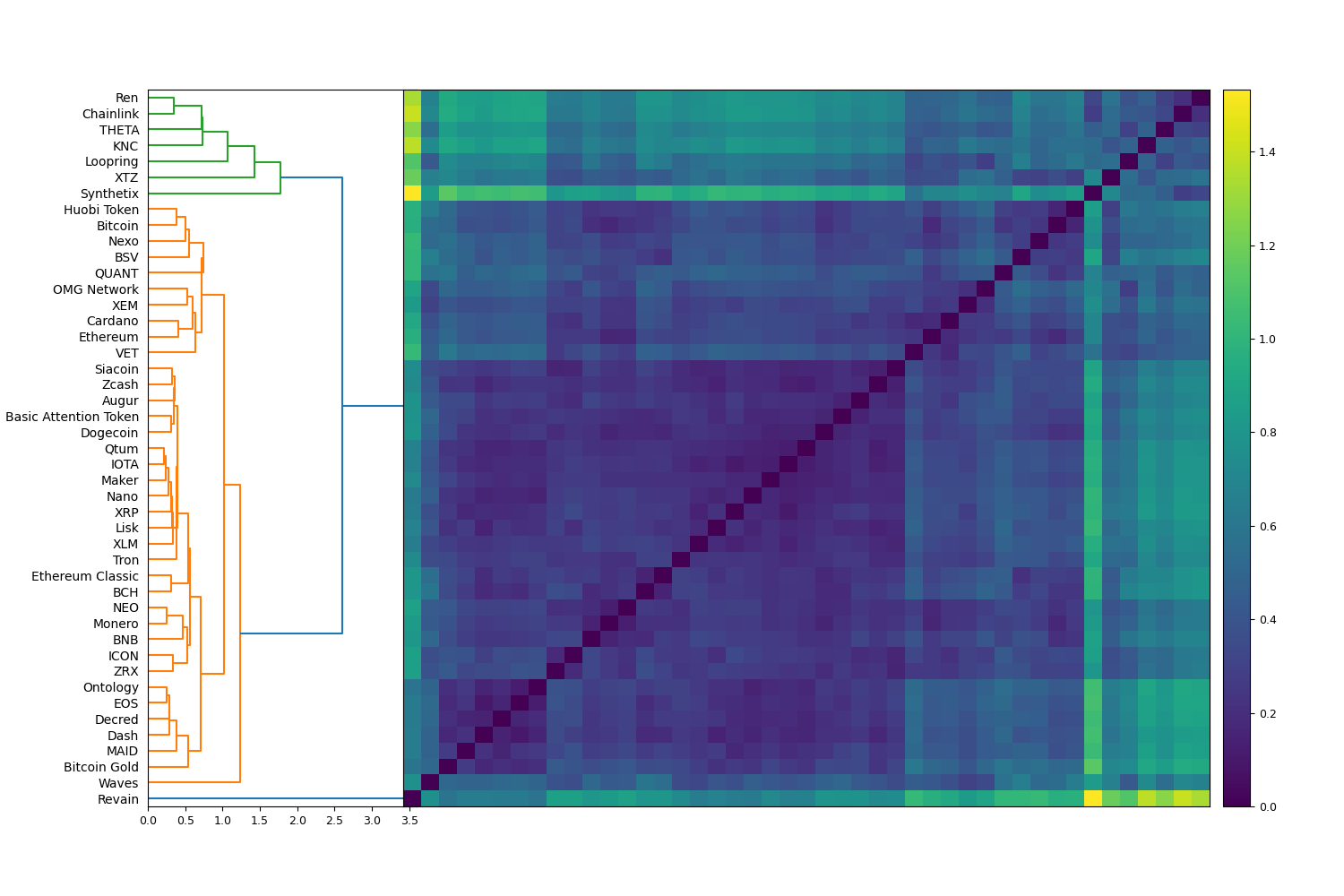}
    \caption{Hierarchical clustering on $D^{TC}$.}
    \label{fig:CryptocurrencyTrajectoryDendrogram}
\end{figure}

\begin{figure}
    \centering
    \includegraphics[width=\textwidth]{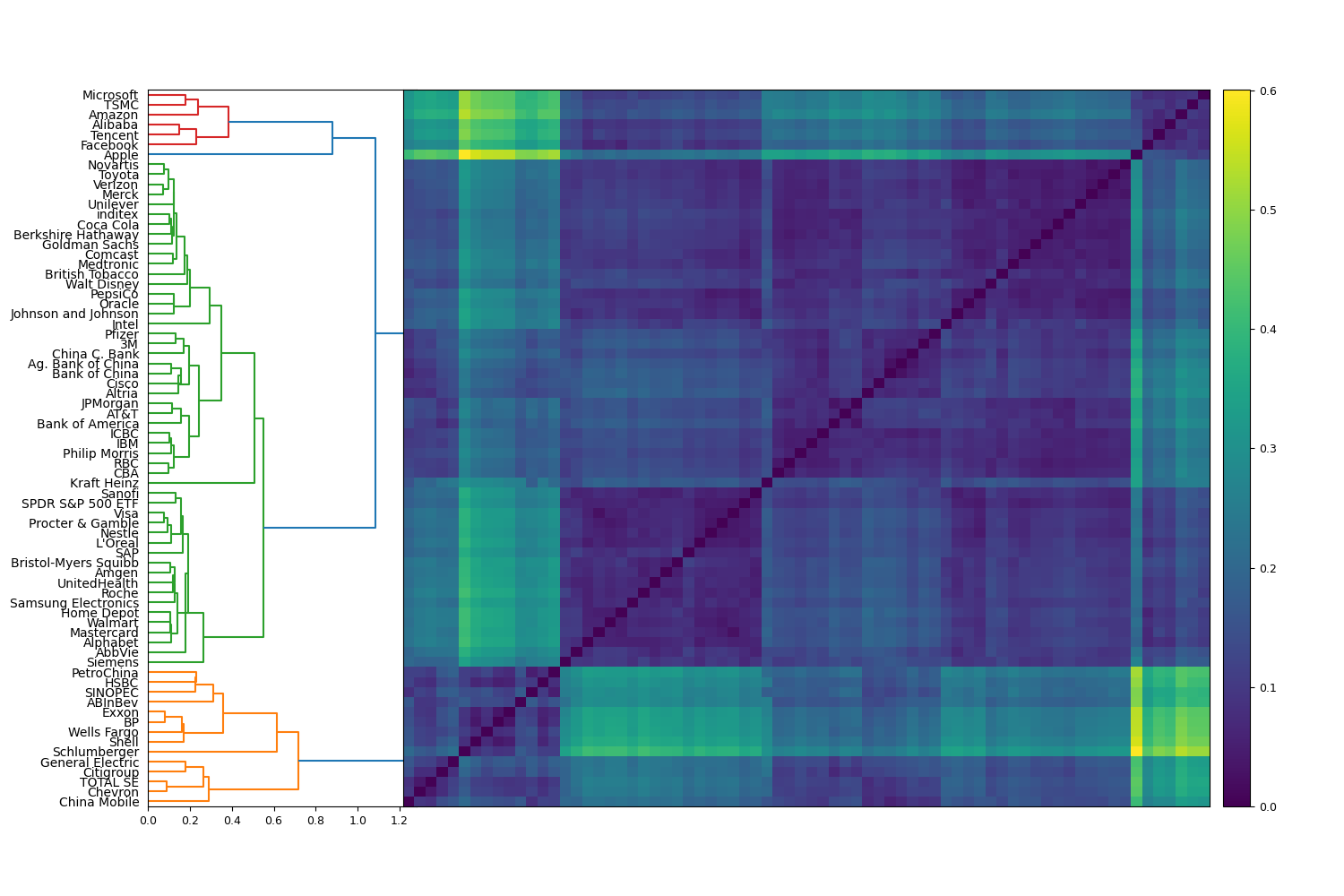}
    \caption{Hierarchical clustering on $D^{TE}$.}
    \label{fig:EquityTrajectoryDendrogram}
\end{figure}

Figures \ref{fig:CryptocurrencyTrajectoryDendrogram} and  \ref{fig:EquityTrajectoryDendrogram} display the cryptocurrency and equity trajectory dendrograms respectively. Each dendrogram highlights several noteable insights regarding trajectory clusters. Figure \ref{fig:CryptocurrencyTrajectoryDendrogram} shows two major clusters, a predominant cluster with high self-similarity, a smaller more amorphous cluster and a single outlier in Revain. The predominant cluster contains most cryptocurrencies analyzed with less volatile price trajectories. The smaller cluster is composed of cryptocurrencies having exhibited major volatility in their price trajectory, such as Chainlink, which experienced a price increase of almost 42 times during our window of analysis. This dendrogram has markedly different structure to the equity trajectory dendrogram, shown in Figure \ref{fig:EquityTrajectoryDendrogram}. The equity trajectory dendrogram exhibits more substantial self-similarity, with three clearly defined clusters; one predominant cluster and two smaller, well-defined clusters exhibiting high self-similarity. The first minority cluster contains stocks such as; Microsoft, Amazon, Alibaba, Tencent, Facebook, Apple and TSMC - all of which are technology companies. The second small cluster contains stocks such as: Chevron, Exxon, BP, Shell, Wells Fargo and HSBC - primarily financial services and energy companies. Both of these sectors tend to perform well in buoyant equity markets, and badly in declining markets. The largest, most predominant third cluster contains the remaining equities. The growth in passive and factor based investing may have increased the similarity in these equities' behaviours, as investors increasingly seek to buy stocks within a sector or `theme' of the market.

\section{Erratic behaviour modelling}
\label{erratic_behaviour}
In this section we study the similarity of structural breaks in our two multivariate time series of log returns, $R^c_i(t)$ and $R^e_j(t)$ defined earlier. For each security in the respective time series, we apply the two-phase \emph{change point detection algorithm} described by \cite{Ross2015} to generate a set of structural breaks for each log return time series. Each change point represents a point in time where the algorithm determines the statistical properties of the time series to have changed. We apply the \emph{Kolmogorov-Smirnov} test, detecting general distributional changes in the underlying time series. The change point detection method could instead focus on changes in specific distributional moments such as mean or variance, however. We obtain two collections of finite sets $\xi^c_1,...,\xi^c_N$, and $\xi^e_1,...,\xi^e_K$ for cryptocurrency and equity time series respectively, where all sets are a subset of  $\{1,...,T\}$.



Next, we compute distances between the cryptocurrency structural break sets $\xi^c_i$ and equity structural break sets $\xi^e_j$. There is significant literature highlighting issues in using metrics such as the Hausdorff distance, due to its sensitivity to outliers, \cite{Baddeley1992, James2020_nsm} and so we use a recently introduced semi-metric modification \cite{James2020_nsm} between candidate sets within our two collections of structural breaks $\xi^c_i$ and $\xi^e_j$. Normalized distances between sets of cryptocurrencies are computed:
\begin{equation}
    D({\xi^c_i},{\xi^c_j}) = \frac{1}{2} \Bigg(\frac{\sum_{b\in \xi^c_j} d(b,\xi^c_i)}{|\xi^c_j|} + \frac{\sum_{a \in {\xi^c_i}} d(a,\xi^c_j)}{|\xi^c_i|} \Bigg),
\end{equation}
where $d(b,\xi^c_i)$ is the minimal distance from $b \in \xi^c_j$ to the set $\xi^c_i$. Distances between sets of equities are computed similarly:
\begin{equation}
    D({\xi^e_i},{\xi^e_j}) = \frac{1}{2} \Bigg(\frac{\sum_{b\in \xi^e_j} d(b,\xi^e_i)}{|\xi^e_j|} + \frac{\sum_{a \in {\xi^e_i}} d(a,\xi^e_j)}{|\xi^e_i|} \Bigg),
\end{equation}
where $d(b,\xi^e_i)$ is the minimal distance from $b \in \xi^e_j$ to the set $\xi^e_i$. This semi-metric is the $L^1$ norm average of all minimal distances between any two sets. As all time series are of equal length, it is not necessary to normalize by the length of the time series. Finally, we form two \emph{breaks matrices} between sets of cryptocurrency structural breaks, $D^{BC}_{ij}=D(\xi^c_i,\xi^c_j)$ and equity structural breaks, $D^{BE}_{ij}=D(\xi^e_i,\xi^e_j)$. To better understand collective similarity in structural breaks, we perform hierarchical clustering on our two breaks matrices.

\begin{figure}
    \centering
    \includegraphics[width=\textwidth]{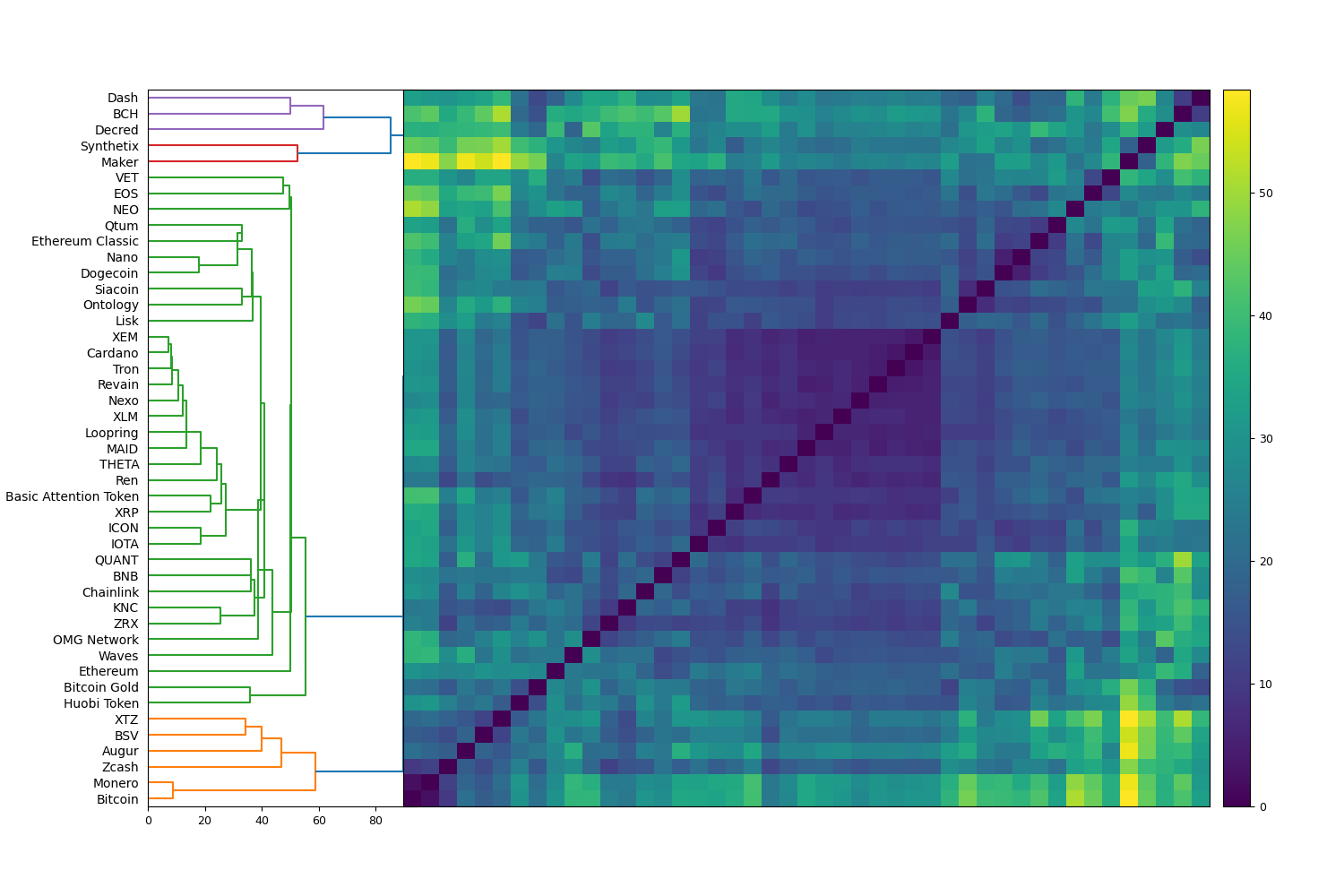}
    \caption{Hierarchical clustering on $D^{BC}$.}
    \label{fig:CryptocurrencyBreaksDendrogram}
\end{figure}

\begin{figure}
    \centering
    \includegraphics[width=\textwidth]{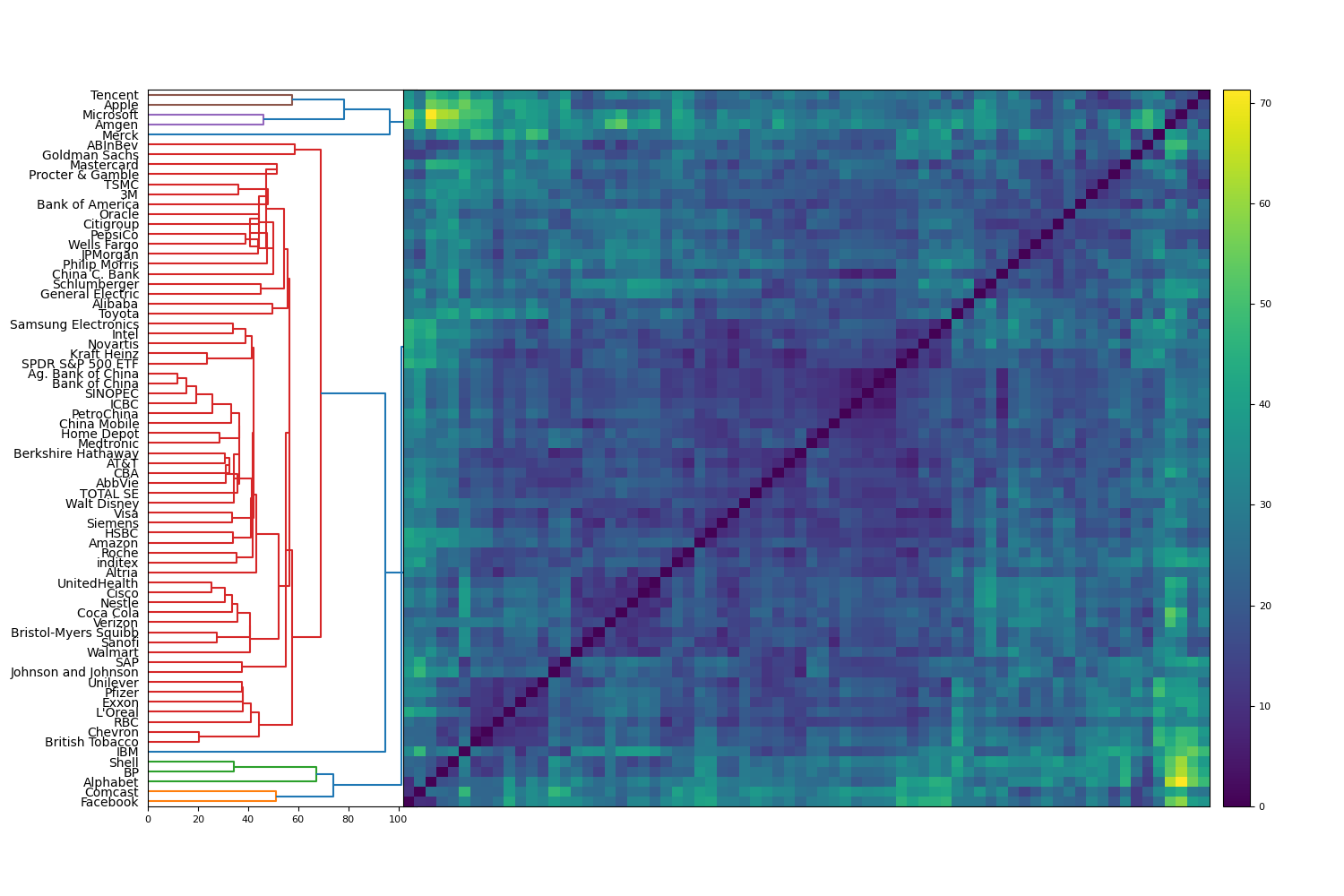}
    \caption{Hierarchical clustering on $D^{BE}$.}
    \label{fig:EquityBreaksDendrogram}
\end{figure}

Like Section \ref{trajectory_modelling}, we compare norms of the two distance matrices to better understand breaks similarity within each collection. As the two collections are of different sizes, again, we normalize the two norm computations by the size of the distance matrix. The normalized cryptocurrency breaks matrix norm $\|D^{BC}\|_{2^{*}} = N^{-1} \sqrt{\sum_{i,j} |d_{ij}^{bc}|^2}$ and the normalized equity breaks matrix norm $\|D^{BE}\|_{2^{*}} = K^{-1} \sqrt{\sum_{i,j} |d_{ij}^{be}|^2}$. The normalized cryptocurrency breaks matrix norm $\|D^{BC}\|_{2^{*}} = 23.42$ and the normalized equity breaks matrix norm $\|D^{BE}\|_{2^{*}} = 23.88$, highlighting approximately equivalent similarity in structural breaks among equities and cryptocurrencies. This result is consistent with earlier findings \cite{James2021_crypto}, which suggest that although one collection of time series may exhibit more volatility and possibly warrant a higher number of structural breaks, the univariate nature of the change point detection algorithm \cite{Ross2015} detects structural breaks relative to the properties of the particular time series. Therefore, although cryptocurrency time series may exhibit more volatility, the behaviour in their collective structural breaks is not necessarily more or less similar than that of the equity collection.


Next, we compare the cluster structures of the two collections. The cryptocurrency breaks dendrogram, seen in Figure \ref{fig:CryptocurrencyBreaksDendrogram} consists of one primary cluster with three, diffuse sub-clusters. The primary cluster has one predominant sub-cluster of concentrated similarity, with the three remaining, smaller clusters having a more indeterminate form. By contrast, the equity breaks dendrogram in Figure \ref{fig:EquityBreaksDendrogram} has a more easily interpreted cluster structure. There are four small clusters, each of which contains two or three equities, and a predominant cluster which is comprised of the remaining equities. The four small clusters appear to cluster based on sector, where cluster one consists of Facebook and Comcast (technology), cluster two consists of BP and Shell (energy), cluster three consists of Merck and Amgen (biotechnology/pharmaceuticals), and cluster four consists of Tencent, Apple and Microsoft (technology). These results suggest that, with the exception of select equities within specific sectors that exhibit similar structural break patterns, most equities have similar structural breaks behaviour.

\section{Extreme behaviour modelling}
\label{Extreme_behaviour}
In this section, we study anomalies with respect to total returns and extreme behaviors within our collections of cryptocurrencies and equities. First, we outline the procedure to measure distance between extreme values of candidate time series. We let $\mu$ be a probability distribution that stores the extreme values of a cryptocurrency time series $c_i(t)$ or equity time series $e_j(t)$. We assume that $\mu$ is a continuous probability measure of the form $\mu=f(x)dx$, where $dx$ is Lebesgue measure, and $f(x)$ is a probability density function that is non-negative everywhere and integrates to $1$. We study the $10$\% and $90$\% points of density, respectively, by the equations

\begin{align}
\int_{-\infty}^l f(x) dx &= 0.1 \\
\int_{u}^\infty f(x) dx &= 0.1
\end{align}
The range $x\leq l$ gives the most extreme 10\% of the distribution on the left side of the distribution, while the range $x\geq u$ gives the 10\% right most extreme values. The restricted function is defined
\begin{align}
g(x)=f(x) \mathbbm{1}_{\{x \leq l\}\cup \{x\geq u\}}=
\begin{cases}
f(x), x \leq l \\
0, l<x<u \\
f(x), x \geq u.
\end{cases}
\end{align}

Next, we construct an associated measure $\nu=g(x) dx$, with $dx$ as Lebesgue measure. We generate $N$ associated probability measures $\mu^c_1,...,\mu^c_N$, and returns measures $\nu^c_1,...,\nu^c_N$ for our cryptocurrency time series. Similarly we generate $K$ probability measures $\mu^e_1,...,\mu^e_K$, and return measure $\nu^e_1,...,\nu^e_K$ for our equity time series. As all restricted measures are of total size 0.2, we use the Wasserstein metric to compute distances between these truncated distributions. Finally, we form a matrix between the distributional extremities of all time series. Let $D^{EC}_{ij}=d^w(\nu^c_i,\nu^c_j)$ be the matrix between the cryptocurrency extreme return distributions, and $D^{EE}_{ij}=d^w(\nu^e_i,\nu^e_j)$ be the matrix between equity extreme return distributions. 

We define total returns for cryptocurrencies $z^c_i = \sum^T_{t=1} R^c_i(t)$ and equities $z^e_j = \sum^T_{t=1} R^e_j(t)$. We now compute \emph{returns matrices} for cryptocurrency returns, $D_{ij}^{RC} = |z_i^c - z_j^c|$ and equity returns $D_{ij}^{RE} = |z_i^e - z_j^e|$. To identify anomalies with respect to returns and extreme values, we transform the four distance matrices into affinity matrices. That is, a candidate \emph{affinity matrix} $A$, is defined as 
\begin{equation}
    A_{ij} = 1 - \frac{D_{ij}}{\max{\{D\}}},
\end{equation}
where $A$ is symmetric, $A_{ii}=1, 0 \leq A_{ij} \leq 1, \forall i,j.$ We now define an \emph{affinity returns matrix} $A^R$ and an \emph{affinity extremes  matrix} $A^E$ both of which are of dimension 117 x 117, which include all 45 cryptocurrencies and 72 equities analyzed in this paper.

\begin{figure*}
    \centering
    \begin{subfigure}[b]{0.48\textwidth}
        \includegraphics[width=\textwidth]{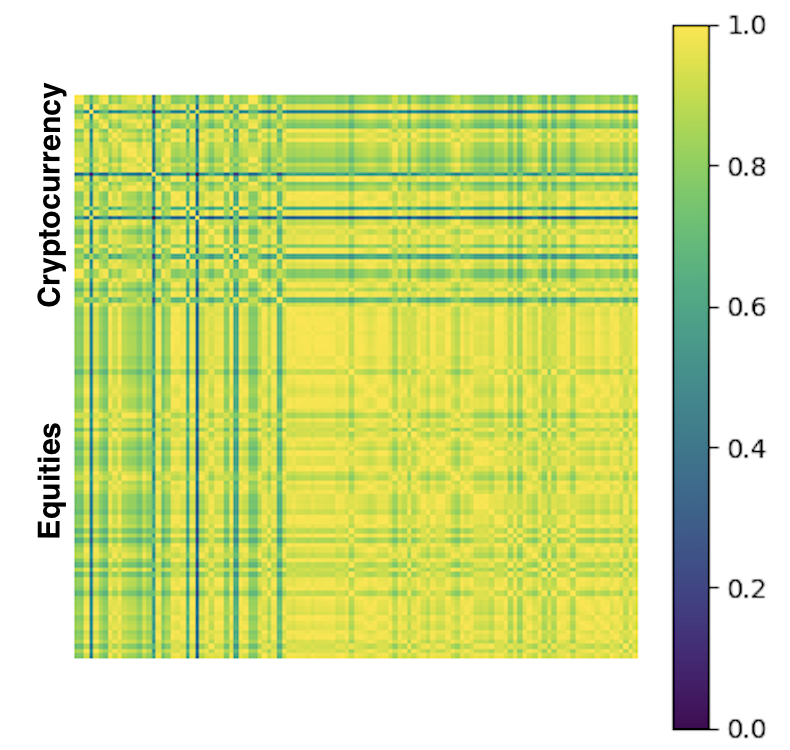}
        \caption{$A^{R}$}
    \label{fig:affinity_asset_returns}
    \end{subfigure}
    \begin{subfigure}[b]{0.48\textwidth}
        \includegraphics[width=\textwidth]{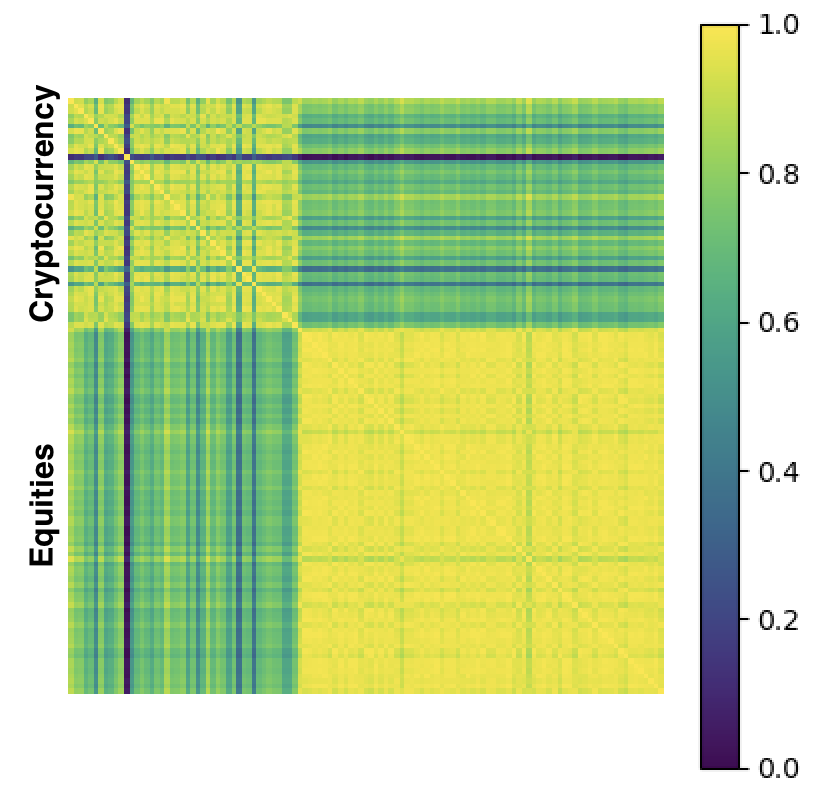}
        \caption{$A^{E}$}
    \label{fig:affinity_asset_extremes}
    \end{subfigure}
        \caption{Affinity matrices returns and extremes.}
    \label{fig:AffinityReturnsExtremes}
\end{figure*}  

We study Figure \ref{fig:AffinityReturnsExtremes} and compare collective similarities in returns and extremes among our two collections. Figure \ref{fig:affinity_asset_returns} shows the affinity returns matrix $A^R$. It is clear that both equities and cryptocurrencies exhibit strong self-similarity, with the equity collection exhibiting slightly more similar returns than the cryptocurrency collection. Although not as strong as the intra-asset similarity, there is still reasonable similarity in the returns profile between that of equities and cryptocurrencies. Figure \ref{fig:affinity_asset_extremes} displays a clear difference in the collective behaviours. Similarly to returns, equities exhibit more self-similarity than that of cryptocurrencies. However unlike returns, there is markedly less similarity when comparing the extremes of cryptocurrencies and equities. These findings are consistent with the results presented in Sections \ref{trajectory_modelling} and \ref{erratic_behaviour}, where cryptocurrencies were shown to be more varied in their trajectories and less consistent in their structural breaks. This finding supports the high degree of dependence identified between extreme and erratic behaviour in the cryptocurrency market.

\begin{figure}
    \centering
    \includegraphics[width=0.75\textwidth]{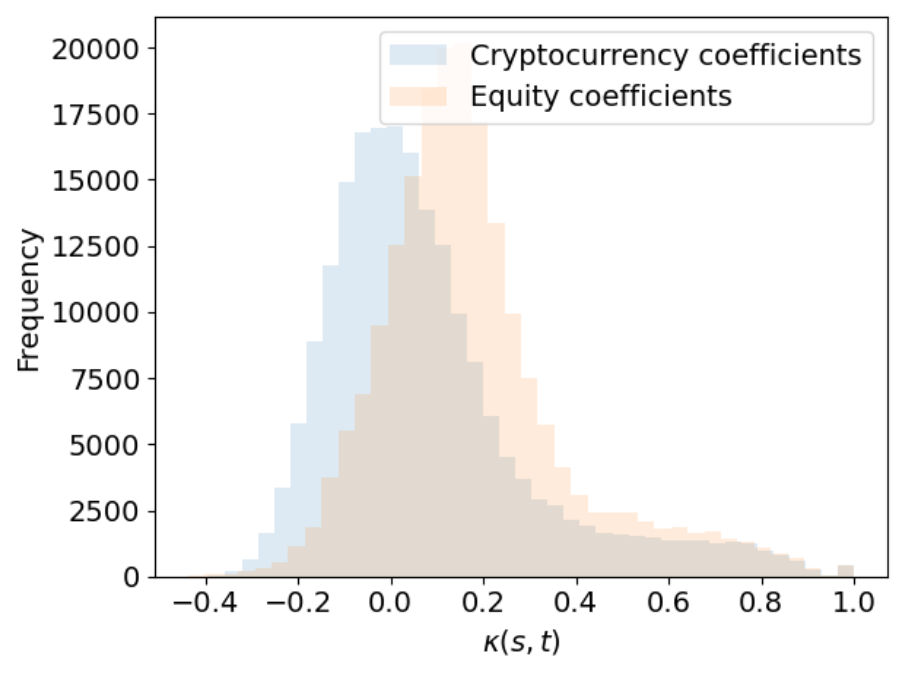}
    \caption{Elements $\kappa^c(s,t)$ and $\kappa^e(s,t)$.}
    \label{fig:TimeCorrelations}
\end{figure}

\vspace{1em}

\section{Anomaly persistence}
\label{anomaly persistence}

 \begin{figure*}
    \centering
    \begin{subfigure}[b]{0.48\textwidth}
        \includegraphics[width=\textwidth]{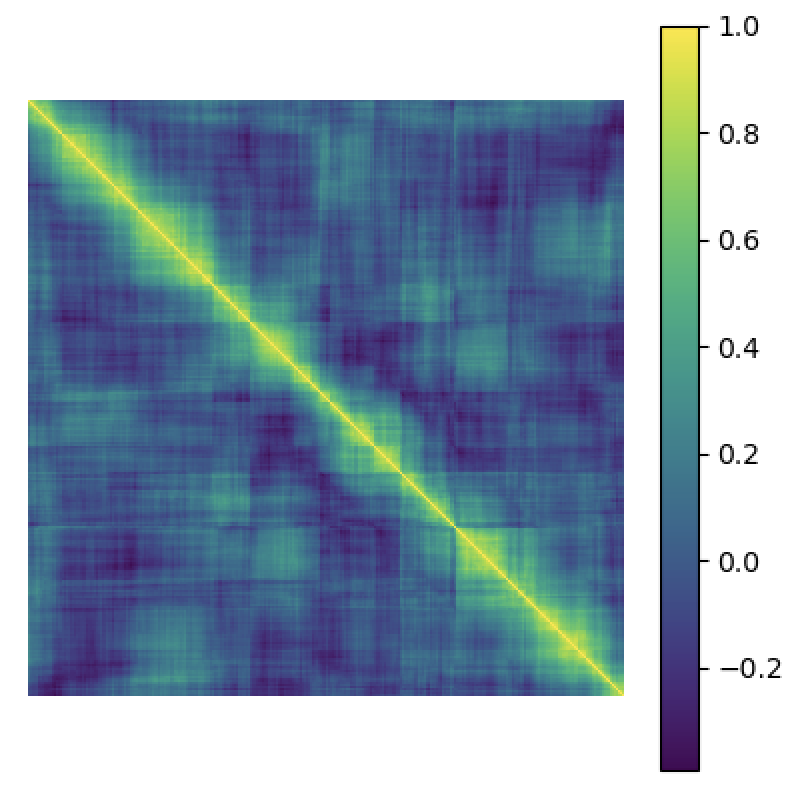}
        \caption{$K^c(s,t)$}
        \label{fig:crypto_time_varying}
    \end{subfigure}
    \begin{subfigure}[b]{0.48\textwidth}
        \includegraphics[width=\textwidth]{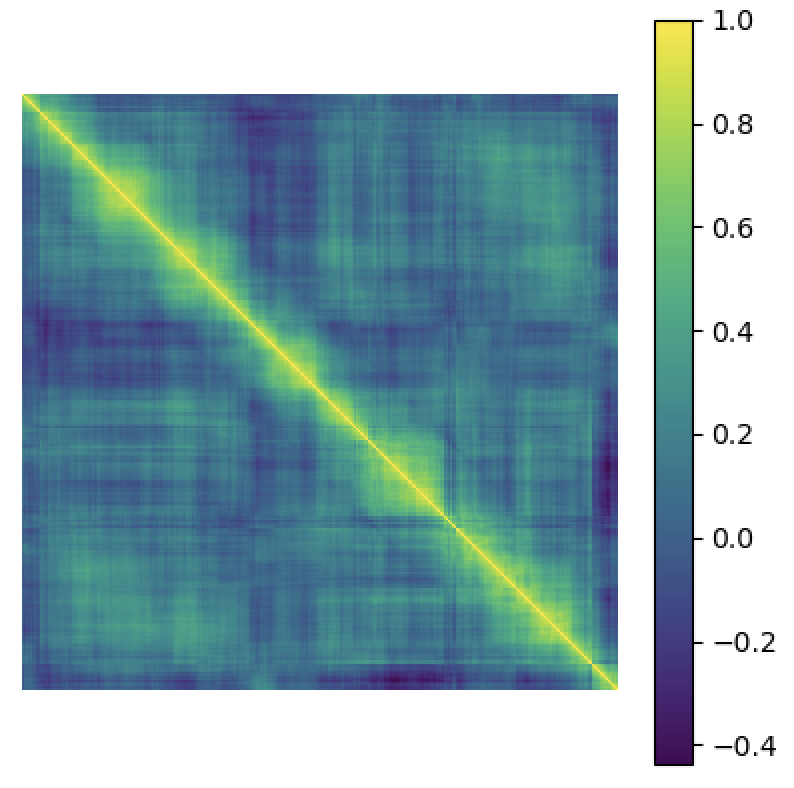}
        \caption{$K^e(s,t)$}
        \label{fig:equity_time_varying}
    \end{subfigure}
    \caption{Anomaly persistence matrices $K^c(s,t)$ and $K^e(s,t)$.}
    \label{fig:KendallTimeVarying}
\end{figure*}

 In this section, we study the evolution of ranks among cryptocurrencies and equities with respect to risk-adjusted returns. Rather than using correlation, we apply the concept of ranks, as financial analysts often apply `ranking' systems when identifying anomalous securities (equities, bonds, currencies, etc.). We define rolling risk-adjusted returns $\kappa^c_i(t) = \frac{\sum^{t}_{m=t-60} R^c_i(m)}{\sigma^c_i(t)}$ and $\kappa^e_j(t) = \frac{\sum^{t}_{m=t-60} R^e_j(m)}{\sigma^e_j(t)}$ where $\sigma^c_i(t)$ and $\sigma^e_j(t)$ are the standard deviations of rolling cryptocurrency and equity log returns for $t \in \{61,...,T\}$. We use a rolling window of 61 days and record two time-evolving sequences of risk adjusted return ranks. Given rank vectors $s,t \in \{61,...,T\}$, we define $K^c(s,t)$ and $K^e(s,t)$ as matrices measuring the Kendall rank correlation coefficient between any two risk-adjusted return rank vectors $s$ and $t$ for all possible points in time. A higher score indicates more similarity in the securities exhibiting high and low risk-adjusted returns at any two points in time. Elements of both matrices $k^c(s,t)$ and $k^e(s,t)$ lie $\in [-1,1]$. 

 \begin{figure*}
    \centering   
    \begin{subfigure}[b]{0.85\textwidth}
        \includegraphics[width=\textwidth]{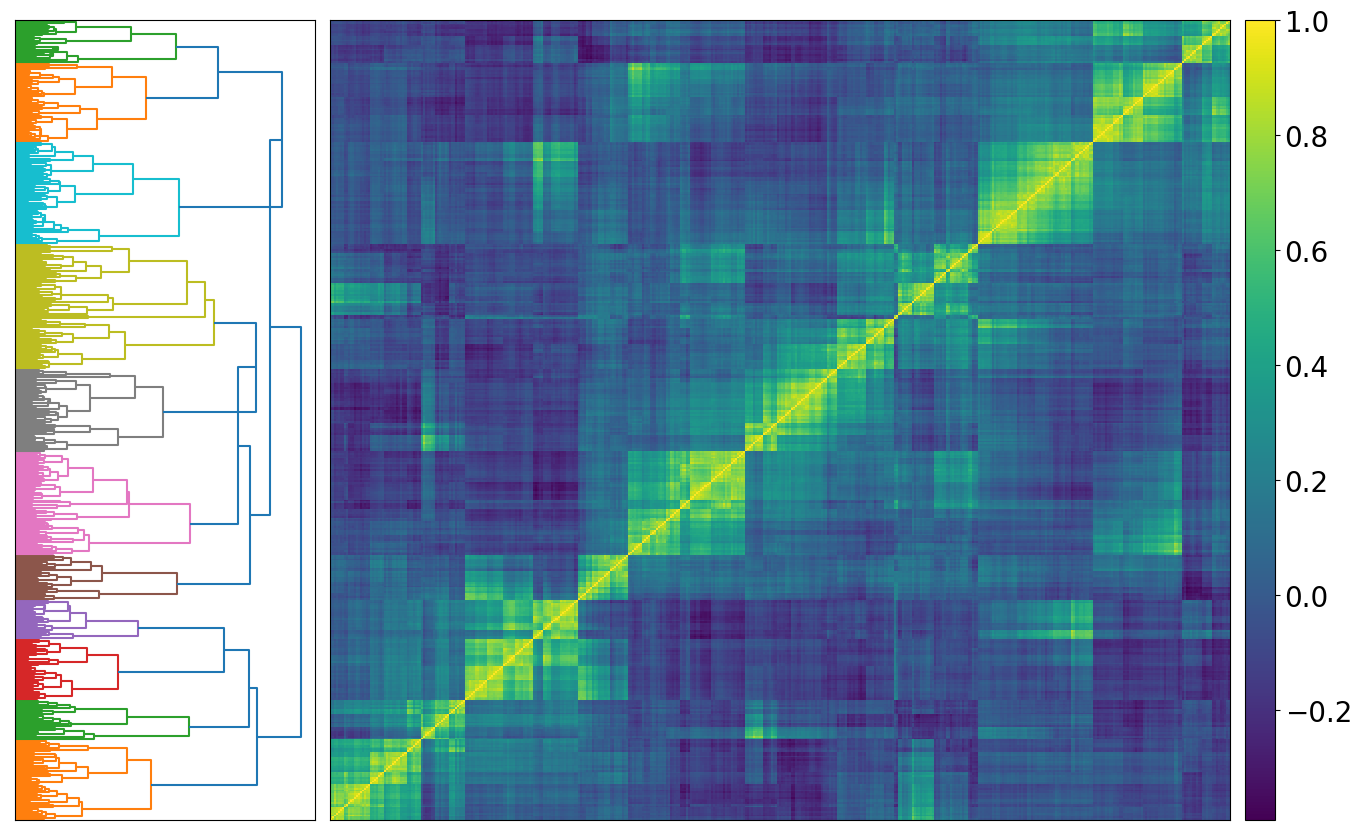}
        \caption{$K^c(s,t)$ }
        \label{fig:kendall_cryptocurrency_dendrogram}
    \end{subfigure}
    \begin{subfigure}[b]{0.85\textwidth}
        \includegraphics[width=\textwidth]{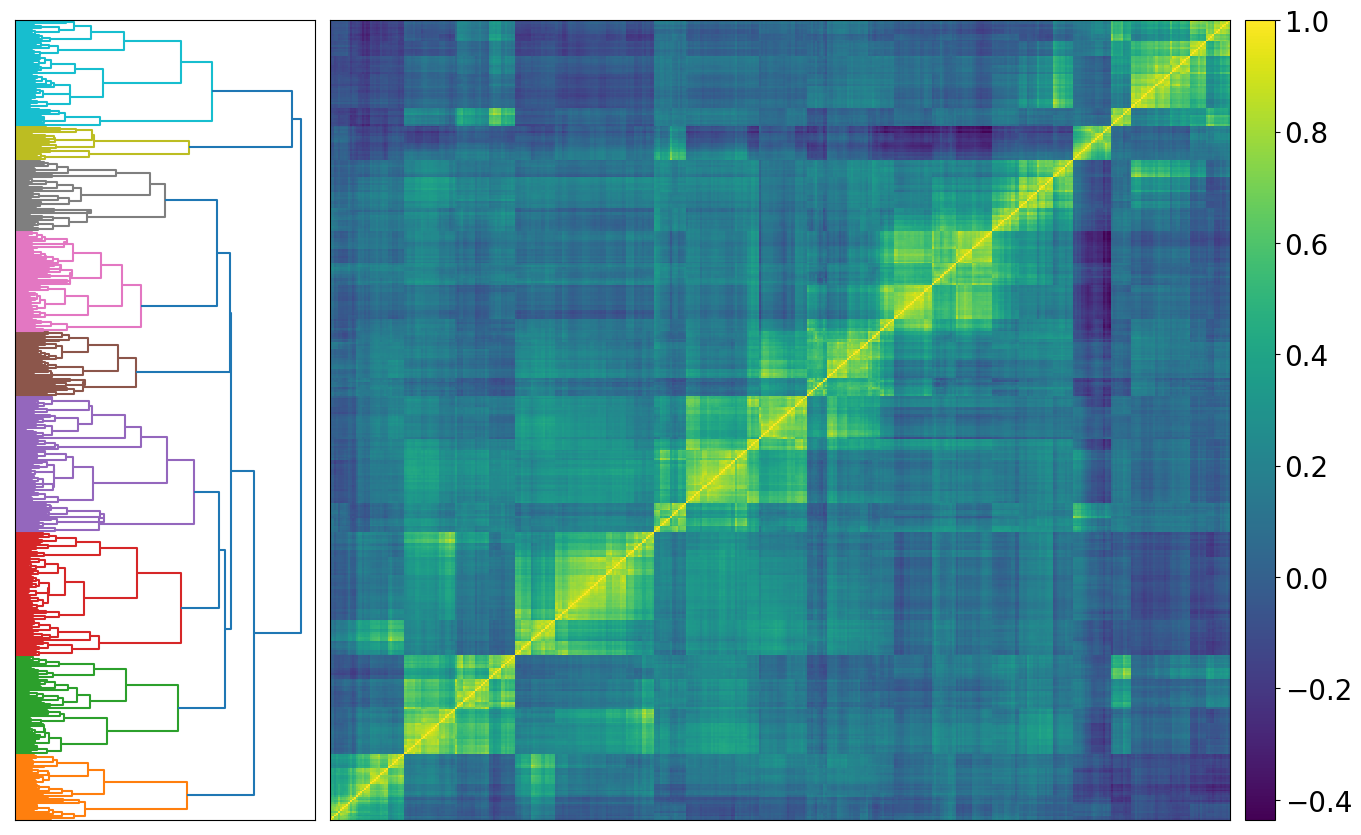}
        \caption{$K^e(s,t)$}
        \label{fig:kendall_equity_dendrogram}
    \end{subfigure}
    \caption{Hierarchical clustering on anomaly persistence matrices $K^c(s,t)$ and $K^e(s,t)$.}
    \label{fig:KendallTimeVaryingDendrogram}
\end{figure*} 

First, we study the norms of the two anomaly persistence matrices $\|K^c\|_2 = \sqrt{\sum_{s,t} |k_{st}^{c}|^2}$ and $\|K^e\|_2 = \sqrt{\sum_{s,t} |k_{st}^{e}|^2}$. The cryptocurrency anomaly persistence norm, $\|K^c\|_2$ = 109.69 and the equity anomaly persistence norm $\|K^e\|_2$ = 122.62. The higher score for the equity collection suggests that there is more consistency in the stocks that are anomalous on a risk-adjusted return basis over time. Figure \ref{fig:TimeCorrelations} which plots two distributions of the elements $k^c(s,t)$ and $k^e(s,t)$, demonstrates a higher average correlation for equities than that of cryptocurrencies. Further interesting structure over time is revealed in Figure \ref{fig:KendallTimeVarying}.

Both Figures \ref{fig:crypto_time_varying} and \ref{fig:equity_time_varying} have high correlation scores around the diagonal, which is indicative of short-term dependence in anomalous behaviours within both collections. However, Figure \ref{fig:equity_time_varying} clearly displays a higher level of similarity over time - indicating more persistence in anomalous behaviours. This is further supported in our hierarchical clustering analysis, where the dendrograms for $K^c(s,t)$ and $K^e(s,t)$ are displayed in Figure \ref{fig:KendallTimeVaryingDendrogram}. There are two primary takeaways from this analysis. First, $K^e(s,t)$ is demonstrably more positive than $K^c(s,t)$ for the vast majority of the dendrogram. This indicates that there is greater similarity in risk-adjusted return ranks, over all comparative measurements in time, for our collection of equities. Second, the $K^e(s,t)$ dendrogram determines a total of 9 clusters, while $K^c(s,t)$ has a total of 11 clusters. This finding indicates more stability in equity rank correlation scores over the entirety of our analysis window.

\newpage
\section{Conclusion}
\label{Conclusion}
Our work in Section \ref{Market_dynamics} demonstrates that collective dynamics in the cryptocurrency market are significantly stronger than that of the equity market. The explanatory variance provided by the largest eigenvector is consistently larger, and more stable among cryptocurrencies than equities. Partitioning our analysis into three discrete windows highlights that collective dynamics are most similar between our two collections during the Peak COVID period, demonstrating that equities and cryptocurrencies behave most similarly during market crises. This is further supported in our correlation matrix analysis, where both cryptocurrencies and equities experience a sharp increase in their correlations during the peak of COVID-19. In periods surrounding the crisis (Pre-COVID and Post-COVID), cryptocurrency correlation coefficients are more strongly positive than that of equities. The findings in this section are also consistent with the work presented in \cite{Drod2020_entropy}. In \cite{Drod2020_entropy} the authors demonstrate that although cryptocurrency dynamics are decoupled with other asset classes during 2019, during select market events in 2020 such as the COVID-19 pandemic, the dynamics of cryptocurrencies and other asset classes behave much more similarly. 

The work presented in this manuscript certainly has its limitations. The work of \cite{Drod2001} demonstrates that after explicitly accounting for time zone differences between indices, (Dow Jones and DAX), there is a significant increase in the similarity of the dynamics of such collections. When initially analyzing both collections in conjunction, the authors show there are two dominant eigenvalues representing the dynamics of each collection. After accounting for differences between the collections by translating the returns of the DAX, one dominant eigenvalue exhibits itself - highlighting a marked increase in the similarity of the dynamics among the total collection. This has not been explicitly considered in this work, and could alter the results and subsequent interpretation. Further research comparing the dynamics of cryptocurrencies and equities using techniques from \cite{Drod2001}, and studying the resulting change in the dynamics deviation scores could be of interest to the econophysics community.

Section \ref{trajectory_modelling} examines the similarity in normalized price trajectories among both collections. Distance matrix norms indicate that equities exhibit more similarity among their trajectories than cryptocurrencies. This may be due to the significant price volatility exhibited by cryptocurrencies over the past two years, making their trajectories (generally, but not universally) more dissimilar. Hierarchical clustering on both time series displays marked differences in cluster structures. Equity trajectories display more self-similarity than cryptocurrencies. We suspect that the latent phenomenon may be the growth of passive and factor-based investing over the last several years.

Section \ref{erratic_behaviour} compares the similarity in erratic behaviour among equities and cryptocurrencies. Distance matrix norms display comparable similarity in the erratic behaviours of cryptocurrencies and equities. Although cryptocurrencies may be more volatile, the univariate nature of our changepoint detection algorithm is unable to determine structural breaks with respect to the rest of the collection. Using other change point detection methodologies to detect structural breaks \cite{Matteson2014} may result in different findings.

Our results in Section \ref{Extreme_behaviour} are consistent with those in Section \ref{trajectory_modelling}. Figure \ref{fig:AffinityReturnsExtremes} shows more homogeneity among equity extremes in comparison to cryptocurrencies. Analyzing distance matrix norms and affinity matrices highlights a substantial difference in self-similarity. When contrasting the similarity in all 117 time series for total returns and extreme values, the distinction in extreme value similarity is most evident.  

Finally in Section \ref{anomaly persistence}, equities are shown to exhibit more persistent anomalies than cryptocurrencies. We apply hierarchical clustering to our proposed anomaly persistence matrix. Hierarchical clustering determines the existence of 9 and 11 clusters respectively in the $K^c(s,t)$ and $K^e(s,t)$. A lower number of clusters signals greater consistency in anomaly ranks over time. This is further supported analyzing the elements of our matrix, which indicate a higher correlation in anomaly rankings over time in the equity time series.

This work bridges several disparate areas of research: nonlinear dynamics, econophysics, COVID-19 and cryptocurrency market dynamics. There are several interesting avenues for future research. First, our analysis could be applied to more asset classes beyond cryptocurrencies and equities. Second, other techniques could be introduced to study phenomena such as: market dynamics, trajectories, extreme and erratic behaviour, and anomaly persistence. Finally, this analysis could be run on different time windows and on a more timely basis. The chaotic and non-deterministic nature of financial markets necessitates timely research on topics of interest.

\section*{Acknowledgements}
I would like to thank Peter Radchenko and Max Menzies for helpful discussions.

\appendix

\section{Mathematical objects glossary}
\label{appendix:mathematical_objects}

\begin{table}[H]
\begin{tabular}{ |p{2.3cm}||p{8.9cm}|}
 \hline
 \multicolumn{2}{|c|}{\textbf{Mathematical objects table: Section \ref{Market_dynamics}}} \\
 \hline
 Object & Description \\
 \hline
 $N$ & \# cryptocurrency time series \\ 
 $K$ & \# equity time series \\
 $c_i(t)$ & Cryptocurrency price time series \\
 $e_j(t)$ & Equity price time series \\
 $R_i^c(t)$ & Cryptocurrency returns time series \\
 $R_j^e(t)$ & Equity returns time series \\
 $\hat{R}^c_i(t)$ & Standardized cryptocurrency returns time series \\
 $\hat{R}^e_j(t)$ & Standardized equity returns time series \\
 $\Omega^c$ & Cryptocurrency correlation matrix \\
 $\Omega^e$ & Equity correlation matrix \\
 $\omega^c(i,j)$ & Element $(i,j)$ in cryptocurrency correlation matrix \\
 $\omega^e(i,j)$ & Element $(i,j)$ in equity correlation matrix \\
 $\Phi^c$ & Cryptocurrency PC coefficient matrix \\
 $\Phi^e$ & Equity PC coefficient matrix \\
 $Z^c$ & Cryptocurrency PC matrix \\
 $Z^e$ & Equity PC matrix \\
 $D^c$ & Cryptocurrency diagonal matrix \\
 $D^e$ & Equity diagonal matrix \\
 $\lambda^c$ & Cryptocurrency eigenvalues \\
 $\lambda^e$ & Equity eigenvalues \\
 $\Lambda^c$ & Cryptocurrency orthogonal eigenvector matrix \\
 $\Lambda^e$ & Equity orthogonal eigenvector matrix \\
 $\Tilde{\lambda}^c$ & Cryptocurrency eigenvalue explanatory variance \\
 $\Tilde{\lambda}^e$ & Equity eigenvalue explanatory variance \\
 $|T_{\text{PRE}}|$ & Length of Pre-COVID period \\
 $|T_{\text{PEAK}}|$ & Length of Peak COVID period \\
 $|T_{\text{POST}}|$ & Length of Post-COVID period \\
 $\text{DD}_{\text{PRE}}$ & Pre-COVID dynamics deviation \\
 $\text{DD}_{\text{PEAK}}$ & Peak COVID dynamics deviation \\
 $\text{DD}_{\text{POST}}$ & Post-COVID dynamics deviation \\
 \hline
\end{tabular}
\caption{Mathematical objects and definitions}
\label{tab:MathematicalObjects1}
\end{table}

\begin{table}[H]
\begin{tabular}{ |p{2.3cm}||p{8.9cm}|}
 \hline
 \multicolumn{2}{|c|}{\textbf{Mathematical objects table: Sections \ref{trajectory_modelling}, \ref{erratic_behaviour}, \ref{Extreme_behaviour}, \ref{anomaly persistence}}} \\
 \hline
 Object & Description \\
 \hline
  $\mathbf{T}_i^{c}$ & Cryptocurrency normalized price trajectory \\
 $\mathbf{T}_j^{e}$ & Equity normalized price trajectory \\
 $D^{TC}$ & Cryptocurrency trajectory matrix \\
 $D^{TE}$ & Equity trajectory matrix \\
 $\|D^{TC}\|_{2^{*}}$ & Normalized cryptocurrency trajectory matrix norm \\
 $\|D^{TE}\|_{2^{*}}$ & Normalized equity trajectory matrix norm \\
 $\xi_1^c,...,\xi_N^c$ & Cryptocurrency structural break sets \\
 $\xi_1^e,...,\xi_K^e$ & Equity structural break sets \\
 $D^{BC}$ & Cryptocurrency breaks matrix \\
 $D^{BE}$ & Equity breaks matrix \\
 $\|D^{BC}\|_{2^{*}}$ & Normalized cryptocurrency breaks matrix norm \\
 $\|D^{BE}\|_{2^{*}}$ & Normalized equity breaks matrix norm \\
 $D^{EC}$ & Cryptocurrency extremes matrix \\
 $D^{EE}$ & Equity extremes matrix \\
 $z_i^c(t)$ & Cryptocurrency total returns time series \\
 $z_j^e(t)$ & Equity total returns time series \\
 $D^{RC}$ & Cryptocurrency returns matrix \\
 $D^{RE}$ & Equity returns matrix \\
 $A^{R}$ & Affinity returns matrix (Cryptocurrencies and equities) \\
 $A^{E}$ &  Affinity extremes matrix (Cryptocurrencies and equities) \\
 $\kappa^c(t)$ & Cryptocurrency risk-adjusted return vector at time $t$\\
 $\kappa^e(t)$ & Equity risk-adjusted return vector at time $t$ \\
 $\sigma_i^c(t)$ & Cryptocurrency realized volatility at time $t$ \\
 $\sigma_j^e(t)$ & Equity realized volatility at time $t$ \\
 $K^c(s,t)$ & Cryptocurrency anomaly persistence matrix \\
 $K^e(s,t)$ & Equity anomaly persistence matrix \\
 $k^c(s,t)$ & Element $(s,t)$ in cryptocurrency anomaly persistence matrix \\
 $k^e(s,t)$ & Element $(s,t)$ in equity anomaly persistence matrix \\
 $\|K^{c}\|_2$ & Cryptocurrency anomaly persistence matrix norm \\
 $\|K^{e}\|_2$ & Equity anomaly persistence matrix norm \\
\hline
\end{tabular}
\caption{Mathematical objects and definitions}
\label{tab:MathematicalObjects2}
\end{table}

\section{Securities analyzed}
\label{appendix:mathematical_objects}

\begin{table}[H]
\begin{tabular}{ |p{1.8cm}|p{3.5cm}|p{1.8cm}|p{3.2cm}|}
 \hline
 \multicolumn{4}{|c|}{\textbf{Cryptocurrency tickers and names}} \\
 \hline
 Ticker & Coin Name & Ticker & Coin Name \\
 \hline
 BTC & Bitcoin & THETA & THETA \\
 ETH & Ethereum & MKR & Maker \\
 XRP & XRP & SNX & Synthetix  \\
 LINK & Chainlink & OMG & OMG Network \\
 BCH & Bitcoin Cash & DOGE & Dogecoin \\
 ADA & Cardano & ONT & Ontology \\
 BNB & Binance Coin & DCR & Decred \\
 XLM & Stellar & BAT & Basic Attention \\
 BSV & Bitcoin SV & NEXO & Nexo \\
 EOS & EOS & ZRX & 0x \\
 XMR & Monero & REN & Ren \\
 TRX & Tron & QTUM & Qtum \\
 XEM & NEM & ICX & ICON \\
 XTZ & Tezos & LRC & Loopring \\
 NEO & NEO Token & KNC & Kyber Network \\
 VET & VeChain & REP & Augur \\
 REV & Revain Classic & LSK & Lisk \\
 DASH & Dash & BTG & Bitcoin Gold \\
 WAVES & Waves & SC & Siacoin \\
 HT & Huobi Token & QNT & QUANT \\
 MIOTA & IOTA & MAID & MaidSafeCoin \\
 ZEC & ZCash & NANO & Nano \\
 ETC & Ethereum Classic &   &   \\ 
\hline
\end{tabular}
\caption{Cryptocurrency tickers and names}
\label{tab:CryptocurrencyTickers}
\end{table}

\begin{table}[H]
{\centering
\begin{tabular}{ |p{3cm}|p{4cm}|p{3.25cm}|p{4cm}|}
 \hline
 \multicolumn{4}{|c|}{\textbf{Equity tickers and names}} \\
 \hline
 Ticker & Equity Name & Ticker & Equity Name \\
 \hline
 NYSE: C & Citigroup & EPA: OR & L'Oreal \\
 NYSE: MRK & Merck & NYSE: UNH & UnitedHealth Group  \\
 NYSE: KO & Coca-Cola & EPA: FP & Total  \\
 NASDAQ: AMGN & Amgen & SHA: 601988 & Bank of China  \\
 NYSE: T & AT\&T & SHA: 601288 & A.B. China \\
 LON: BATS & British American Tobacco & LON: HSBA & HSBC  \\
 NYSE: JPM & JP Morgan Chase & NYSE: VZ & Verizon \\
 TYO: 7203 & JP Toyota Motor & NYSEARCA: SPY & SPDR S\&P 500 \\
 ASX: CBA & CBA & NYSE: SLB & Schlumberger \\
 SHA: 601939 & China Construction Bank & EPA: SAN & Sanofi \\
 NASDAQ: CSCO & Cisco  & NYSE: IBM & IBM \\
 NYSE: MDT & Medtronic  & NYSE: PG & Procter \& Gamble\\
 LON: BP & BP  & NASDAQ: FB & Facebook \\
 NYSE: BRK & Berkshire Hathaway  & SHA: 601398 & ICBC \\
 SWX: NOVN & Novartis  & SHA: 600028 & Sinopec \\
 ETR: SIE & Siemens  & NASDAQ: MSFT & Microsoft \\
 NYSE: WMT & Walmart & NYSE: WFC & Wells Fargo  \\
 NYSE: DIS & Walt Disney  & SWX: RO & Roche Holdings \\
 NYSE: JNJ & Johnson and Johnson & NASDAQ: PEP & PepsiCo \\
 NASDAQ: INTC & Intel & NYSE: PFE & Pfizer \\
 NYSE: GE & General Electric & NYSE: XOM & Exxon Mobil \\
 NASDAQ: AAPL & Apple  & NYSE: BMY & Bristol-Myers Squibb \\
 NYSE: GS & Goldman Sachs & NASDAQ: CMCSA & Comcast \\
 KRX: 005930 & Samsung Electronics & NYSE: HD & Home Depot  \\
 HKG: 0941 & China Mobile & NYSE: MA & Mastercard \\
 NASDAQ: AMZN & Amazon & LON: ULVR & Unilver  \\
 NYSE: ORCL & Oracle & SWX: NESN & Nestle \\
 NYSE: MMM & 3M & NYSE: V & Visa  \\
 AMS: RDSA & Royal Dutch Shell & NYSE: PM & Philip Morris\\
 BME: ITX & INDITEX & NYSE: ABBV & AbbVie Inc \\
 NYSE: MO & Altria Group  & HKG: 9988 & Alibaba \\
 NYSE: CVX & Chevron  & NASDAQ: KHC & Kraft Heinz \\
 TSE: RY & Royal Bank of Canada & NASDAQ: GOOGL & Alphabet   \\
 HKG: 0700 & Tencent  & EBR: ABI & Anheuser Busch Inbev NV \\
 TPE: 2330 & TSMC & NYSE: BAC & Bank of America \\
 ETR: SAP & SAP & SHA: 601857 & PetroChina\\ 
\hline
\end{tabular}}
\caption{Equity tickers and names}
\label{tab:EquityTickers}
\end{table}

\section{Change point detection algorithm}
\label{appendix:CPD}
In this section, we provide an outline of change point detection algorithms, and describe the specific algorithm that we implement. Many statistical modelling problems require the identification of \emph{change points} in sequential data. By definition, these are points in time at which the statistical properties of a time series change. The general setup for this problem is the following: a sequence of observations $x_1,x_2,...,x_n$ are drawn from random variables $X_1, X_2,...,X_n$ and undergo an unknown number of changes in distribution at points $\tau_1,...,\tau_m$. One assumes observations are independent and identically distributed between change points, that is, between each change points a random sampling of the distribution is occurring. Following Ross \cite{Ross2015}, we notate this as follows:
\begin{equation*}
    X_{i} \sim 
    \begin{cases}
      F_{0} \text{ if } i \leq \tau_1 \\
      F_{1} \text{ if } \tau_1 < i \leq  \tau_2  \\
      F_{2} \text{ if } \tau_2 < i  \leq \tau_3,  \\
      \hdots
    \end{cases}
\end{equation*}
While this requirement of independence may appear restrictive, dependence can generally be accounted for by modelling the underlying dynamics or drift process, then applying a change point algorithm to the model residuals or one-step-ahead prediction errors, as described by Gustafsson \cite{gustafsson2001}. The change point models applied in this paper follow Ross \cite{Ross2015}.

\subsection{Batch change detection (Phase I)}
This phase of change point detection is retrospective. We are given a fixed length sequence of observations $x_1,\ldots,x_n$ from random variables $X_1,\ldots,X_n$. For simplicity, assume at most one change point exists. If a change point exists at time $k$, observations have a distribution of $F_0$ prior to the change point, and a distribution of $F_1$ proceeding the change point, where $F_0 \neq F_1$. That is, one must test between the following two hypotheses for each $k$: 

\begin{equation*}
    H_0: X_{i} \sim F_0, i = 1,...,n
\end{equation*}

\begin{equation*}
    H_1: X_{i} \sim 
    \begin{cases}
      F_{0} & i = 1,2,...,k \\
      F_{1}, & i = k + 1, k+2, ..., n  \\
    \end{cases}
\end{equation*}
and end with the choice of the most suitable $k$.

One proceeds with a two-sample hypothesis test, where the choice of test is dependent on the assumptions about the underlying distributions. To avoid distributional assumptions, non-parametric tests can be used. Then one appropriately chooses a two-sample test statistic $D_{k,n}$ and a threshold $h_{k,n}$. If $D_{k,n}>h_{k,n}$ then the null hypothesis is rejected and we provisionally assume that a change point has occurred after $x_k$. These test statistics $D_{k,n}$ are normalised to have mean $0$ and variance $1$ and evaluated at all values $1 < k < n$, and the largest value is assumed to be coincident with the existence of our sole change point. That is, the test statistic is then

\begin{equation*}
    D_{n} = \max_{k=2,...,n-1} D_{k,n} = \max_{k=2,...,n-1} \Bigg| \frac{\Tilde{D}_{k,n} - \mu_{\Tilde{D}_{k,n}}}{\sigma_{\Tilde{D}_{k,n}}}  \Bigg|
\end{equation*}
where $\Tilde{D}_{k,n}$ were our unnormalised statistics. 

The null hypothesis of no change is then rejected if $D_{n} > h_n$ for some appropriately chosen
threshold $h_n$. In this circumstance, we conclude that a (unique) change point has occurred and its location is the value of $k$ which maximises $D_{k,n}$. That is,

\begin{equation*}
    \hat{\tau} = \argmax_k D_{k,n}.
\end{equation*}

This threshold $h_n$ is chosen to bound the Type 1 error rate as is standard in statistical hypothesis testing. First, one specifies an acceptable level $\alpha$ for the proportion of false positives, that is, the probability of falsely declaring that a change has occurred if in fact no
change has occurred. Then, $h_n$ should be chosen as the upper $\alpha$ quantile of the distribution
of $D_n$ under the null hypothesis. For the details of computation of this distribution, see \cite{Ross2015}. Computation can often be made easier by taking appropriate choice and storage of the $D_{k,n}$.

\subsection{Sequential change detection (Phase II)}
In this case, the sequence $(x_t)_{t \geq 1}$ does not have a fixed length. New observations are received over time, and multiple change points may be present. Assuming no change point exists so far, this approach treats $x_1, . . . , x_t$ as a fixed length sequence and computes $D_t$ as in phase I. A change is then flagged if $D_t > h_t$ for some appropriately chosen threshold. If no change is detected, the next observation $x_{t+1}$ is brought into the sequence. If a change is detected, the process restarts from the following observation in the sequence. The procedure therefore consists of a repeated sequence of hypothesis tests.

In this sequential setting, $h_t$ is chosen so that the probability of incurring a Type 1 error is constant over time, so that under the null hypothesis of no change, the following holds:

\begin{equation*}
    P(D_1 > h_1) = \alpha,
\end{equation*}

\begin{equation*}
    P(D_t > h_t | D_{t-1} \leq h_{t-1}, ... , D_{1} \leq h_{1}) = \alpha, \ t > 1.
\end{equation*}
In this case, assuming that no change occurs, the average number of observations received before a false positive detection occurs is equal to $\frac{1}{\alpha}$. This quantity is referred to as the average run length, or ARL0. Once again, there are computational difficulties with this conditional distribution and the appropriate values of $h_t$, as detailed in Ross \cite{Ross2015}.

\clearpage
\bibliographystyle{elsarticle-num-names}
\bibliography{References.bib}
\end{document}